\documentclass{aa}
\usepackage[varg]{txfonts} 

\usepackage{graphicx}
\begin{document}

   \title{Cepheid Metallicity in the Leavitt Law (C-MetaLL) Survey\\  }
   \subtitle{V. New multiband ($grizJHK_s$) Cepheid light curves and period--luminosity relations}

   \author{A. Bhardwaj\fnmsep\thanks{Marie Skłodowska-Curie Fellow}
          \inst{1}
         \and
          V. Ripepi\inst{1}        
           \and
          V. Testa\inst{2}         
          \and
          R. Molinaro\inst{1} %
          \and
          M. Marconi\inst{1}
        \and
          G. De Somma\inst{1,3}
        \and
          E. Trentin\inst{1,4,5}
        \and
          I. Musella\inst{1}
        \and
          J. Storm\inst{4}
        \and
          T. Sicignano\inst{6,1} 
        \and
          G. Catanzaro\inst{7}
          }

        \institute{INAF-Osservatorio Astronomico di Capodimonte, Salita Moiariello 16, 80131, Naples, Italy\\
        \email{anupam.bhardwajj@gmail.com; anupam.bhardwaj@inaf.it}
        \and
        INAF-Osservatorio Astronomico di Roma, Via Frascati 33, 00078, Monte Porzio Catone, Italy
 \and
 Istituto Nazionale di Fisica Nucleare (INFN)-Sez. di Napoli, Compl. Univ.di Monte S. Angelo, Edificio G, Via Cinthia, 80126 Napoli, Italy
 \and
 Leibniz-Institut für Astrophysik Potsdam (AIP), An der Sternwarte 16, D-14482 Potsdam, Germany
 \and
 Institut für Physik und Astronomie, Universität Potsdam, Haus 28, Karl-Liebknecht-Str. 24/25, D-14476 Golm (Potsdam), Germany
 \and 
 Scuola Superiore Meridionale, Largo S. Marcellino 10, 80138 Napoli, Italy
 \and
 INAF-Osservatorio Astrofisico di Catania, Via S. Sofia 78, 95123, Catania, Italy
        }

   \date{Received December 21, 2023; accepted December 22, 2023}

 
  \abstract
   {The highly debated effect of metallicity on the absolute magnitudes of classical Cepheid variables needs to be properly quantified for determining accurate and precise distances based on their Leavitt Law.
   }
   {Our goal is to obtain homogeneous optical and near-infrared light curves of Milky Way Cepheid variables complementing their already collected high-resolution spectroscopic metallicities as part of the
   C-MetaLL survey. Together with {\it Gaia} parallaxes, we   investigate period-luminosity-metallicity relations for Cepheid variables at multiple wavelengths.
   }
   {We present homogeneous multiband ($grizJHK_s$) time-series observations of 78 Cepheids including 49 fundamental mode variables and 29 first-overtone mode variables. These observations were collected
   simultaneously using the ROS2 and REMIR instruments at the Rapid Eye Mount telescope. Multiwavelength photometric data were used to investigate pulsation properties of Cepheid variables and 
   derive their period--luminosity (PL) and period--Wesenheit (PW) relations. 
    }
   {The Cepheid sample covers a large range of distances ($0.5-19.7$~kpc) with varying precision of parallaxes, and thus astrometry-based luminosity fits were used to derive PL and PW relations 
   in optical Sloan ($griz$) and near-infrared ($JHK_s$) filters. These empirically calibrated relations exhibit large scatter primarily due to larger uncertainties in parallaxes of distant 
   Cepheids, but their slopes agree well with those previously determined in the literature. Using homogeneous high-resolution spectroscopic metallicities of 61 Cepheids covering 
   $-1.1<\textrm{[Fe/H]}<0.6$~dex, we quantified the metallicity dependence of PL and PW relations which varies between $-0.30\pm0.11$ (in $K_s$) and $-0.55\pm0.12$ (in $z$) mag/dex in 
   $grizJHK_s$ bands. However, the metallicity dependence in the residuals of the PL and PW relations is predominantly seen for metal-poor stars ($\textrm{[Fe/H]}<-0.3$~dex), which also 
   have larger parallax uncertainties. The modest sample size precludes us from separating the contribution to the residuals due to parallax uncertainties, metallicity effects, and reddening 
   errors. While this Cepheid sample is not optimal for calibrating the Leavitt law, upcoming photometric and spectroscopic datasets of the C-MetaLL survey will allow the accurate derivation of PL 
   and PW relations in the Sloan and near-infrared bandpasses, which will be useful for the distance measurements in the era of the  Vera C. Rubin Observatory's Legacy Survey of Space and Time and 
   upcoming extremely large telescopes.  
   }
   {}
   

\titlerunning{Multiband light curves of Cepheid variables and period-luminosity-metallicity relations}
\authorrunning{Bhardwaj A. et al.}

   \maketitle
%

\section{Introduction}

Classical Cepheid variables in the Milky Way (MW) provide the absolute calibration of their period--luminosity (PL) relation or the  Leavitt law \citep{leavitt1912} based on their individual distances derived from geometric parallaxes. The absolute calibration of the  Leavitt law is crucial in order  to measure extragalactic distances and determine the present expansion rate of the Universe through the  `cosmic distance ladder' \citep{freedman2001, riess2022}. The present expansion rate in the late evolutionary universe,   the local Hubble constant, is currently in discord with its early universe measurement from the {\it Planck} mission \citep{riess2022, planck2020}. This tension between the Hubble constant measurements from the two extreme ends of the Universe may hint at new or missing physics in the standard cosmological model \citep{valentino2021, abdalla2022}. 

The accuracy and the precision of Cepheid-based distance measurements in the traditional cosmic distance ladder are now only limited by the systematic uncertainties in the calibration of their PL relation and its metallicity dependence \citep[][]{riess2022, bhardwaj2023a}. Before {\it Gaia}, the absolute calibration of Cepheid PL  relations was mostly based on a limited sample of Cepheids with {\it Hubble Space Telescope} parallaxes \citep{benedict2007, riess2014, riess2018} or using distances determined from independent methods, such as the Baade-Wesselink or infrared surface brightness methods \citep{gieren1998, fouque2007, storm2011, bhardwaj2016a}. This is now changing thanks to increasingly more accurate and precise geometric parallaxes of thousands of Cepheids in the MW from the {\it Gaia} mission \citep{prusti2016, vallenari2023, ripepi2022}. Several recent studies have utilized unprecedentedly precise {\it Gaia} parallaxes to calibrate PL relations at multiple wavelengths  \citep[e.g.][]{groenewegen2018, ripepi2021, ripepi2022, breuval2022, riess2022a, reyes2023, narloch2023}. It has been claimed that some of these calibrations of Cepheid luminosities have reached percent-level precision at specific wavelengths \citep{riess2022a,reyes2023}.

The increasing precision of {\it Gaia} parallaxes also enabled some of the above-mentioned studies to investigate the metallicity dependence of Cepheid PL relations at multiple wavelengths. However, there is currently no consensus on the metallicity dependence of Cepheid period-luminosity-metallicity (PLZ) and period-Wesenheit-metallicity (PWZ) relations at multiple wavelengths. Most of the recent empirical studies have found a negative metallicity coefficient with values between $-0.2$ and $-0.5$ mag/dex \citep[e.g. Figure 19 in][]{trentin2023a}, which was also confirmed by models for the Wesenheit relations \citep{anderson2016, desomma2022}. Nevertheless, these coefficients are often either weakly constrained and/or differ by a factor of two among recent empirical studies \citep[e.g.][]{ripepi2021, riess2021, riess2022, breuval2022, molinaro2023, bhardwaj2023a, trentin2023a}. For example, \citet{trentin2023a} found a metallicity coefficient of $-0.458\pm0.052$~mag/dex in $K_s$-band for the MW Cepheid PL relation, while \citet{breuval2022} found a coefficient of $-0.321\pm0.068$~mag/dex using Cepheids in the Galaxy and the Magellanic Clouds. Recently, \citet{bhardwaj2023a} determined a metallicity term of $-0.43\pm0.18$~mag/dex in $K_s$-band using individual metallicities of MW Cepheids. This coefficient is better constrained ($-0.33\pm0.07$~mag/dex) using MW and Large Magellanic Cloud (LMC) Cepheids within a very narrow range of their mean metallicities ($\Delta$[Fe/H]$=0.46$~dex). Homogeneous high-resolution spectroscopic metallicities of MW Cepheids having a wide range of metallicity and multiband photometry are needed to complement their {\it Gaia} parallaxes to better constrain the zero-points and the metallicity coefficients of Cepheid PLZ relations. 

Most of the recent studies on the empirical calibration of Cepheid PL or PLZ relations in $BVIJHK_s$ bands utilized heterogeneous optical and infrared photometry together with homogeneous {\it Gaia} data. The optical PL relations, except those in the {\it Gaia} photometric systems, primarily used one of the largest compilations of photoelectric observations of 894 Cepheids in standard Johnson-Cousin-Kron filters \citep[$UBV(RI)c$,][]{berdi2008} or the modern CCD photometry of Cepheids in $BVI$ bands \citep[e.g.][]{berdi2015}. Time-domain variability surveys such as the Optical Gravitational Lensing Experiment \citep[OGLE,][]{udalski2018} and the All-Sky Automated Survey for Supernovae \citep[ASAS-SN,][]{jayasinghe2018} have also provided optical $V$ and/or $I$ light curves of thousands of Cepheids in the MW, but lack   multiwavelength coverage. Moreover, several ongoing large-scale variability surveys are now being carried out in the Sloan ($ugriz$) photometric system such as the Zwicky Transient Facility \citep[ZTF,][in $gri$]{bellm2019}, enabling the discovery and identification of hundreds of Cepheids in the MW \citep{chen2020}.

However, the pulsation properties of Cepheid variables and their PL relations have not yet been explored   in detail in the specific Sloan filters. \citet{hoffman2015} used the  8.1m Gemini North telescope to observe Cepheids and derive their PL relations in the  $gri$ filters in NGC 4258, which is one of the anchor galaxies for the extragalactic distance scale \citep{riess2022}. \citet{kodric2018} presented PL relations for Cepheids in the Andromeda galaxy in Sloan-like ($gri$) filters using the data from the Panoramic Survey Telescope And Rapid Response System survey \citep[][]{tory2012}. Recently, \citet{adair2023} presented the largest sample of more than 1600 Cepheids in M33 using photometry in the $gri$ bands. \citet{narloch2023} provided light curves of 96 MW Cepheids and their PL relations in the Sloan ($gri$) filters. Similar calibrations of $gri$ band PL relations for RR Lyrae, Type II Cepheids, and the  anomalous Cepheids in the globular clusters have also been provided using the ZTF data \citep{ngeow2022, ngeow2022a, ngeow2022b}. In view of the much anticipated Vera C. Rubin Observatory's Legacy Survey of Space and Time \citep[LSST,][]{ivezic2019}, which will operate at $ugrizy$ wavelengths, it is crucial to provide an empirical calibration of PL relations for pulsating stars in Sloan filters. 

Homogeneous near-infrared (NIR) $JHK_s$ band light curves of MW Cepheids for the largest sample of 131 Cepheids was provided by \citet{monson2011}. The empirical calibrations of Cepheid PL relations in $JHK_s$ bands in the literature are either based on the above-mentioned dataset combined with older photometric measurements \citep[e.g.][]{welch1984, laney1994, barnes1997} or using random phase-corrected Two Micron All Sky Survey (2MASS) observations \citep{skrutskie2006}. NIR photometry offers several advantages over optical bands, due to lower extinction and  smaller temperature variations leading to less scatter in the PL relations \citep[see the review by][]{bhardwaj2020}. However, smaller variability amplitudes and near-sinusoidal light curves of Cepheid variables in the NIR bands can complicate their identification and classification, which is easier in the optical domain due to asymmetric features and larger amplitudes \citep{bhardwaj2015, bhardwaj2017, udalski2018}. Given that all upcoming large observational facilities will mostly operate at infrared wavelengths, it is imperative to obtain homogeneous optical and infrared photometric light curves of Cepheids in the MW to fully utilize current and future {\it Gaia} parallaxes for calibrating the Leavitt law at multiple wavelengths.  

This paper presents simultaneous optical ($griz$) and NIR ($JHK_s$) light curves of 78 classical Cepheids for the first time. The manuscript is the fifth in the series of the Cepheid Metallicity in the Leavitt Law (C-MetaLL) survey, which aims to provide homogeneous time-series photometry and high-resolution spectroscopy of Cepheids in our Galaxy \citep{ripepi2021, trentin2023}. While previous C-MetaLL publications explored the impact of new spectroscopic observations of Cepheids on their PLZ relations, this paper   presents the first results of the ongoing photometric programme. The details regarding the photometric observations, the data reduction and analysis, and the photometric calibration are presented in Section~\ref{sec:data}. Multiwavelength light curves of Cepheids and their PL relations are discussed in Sections~\ref{sec:var} and \ref{sec:plrs}, respectively. The impact of metallicity on the Leavitt law is discussed in Section~\ref{sec:plz}, and the results of this work are summarized in Section~\ref{sec:discuss}.
\section{Data and photometry} \label{sec:data}

\subsection{The sample of Cepheid variables}
\label{sec:data_sample}

\begin{figure}
  \includegraphics[width=0.98\columnwidth]{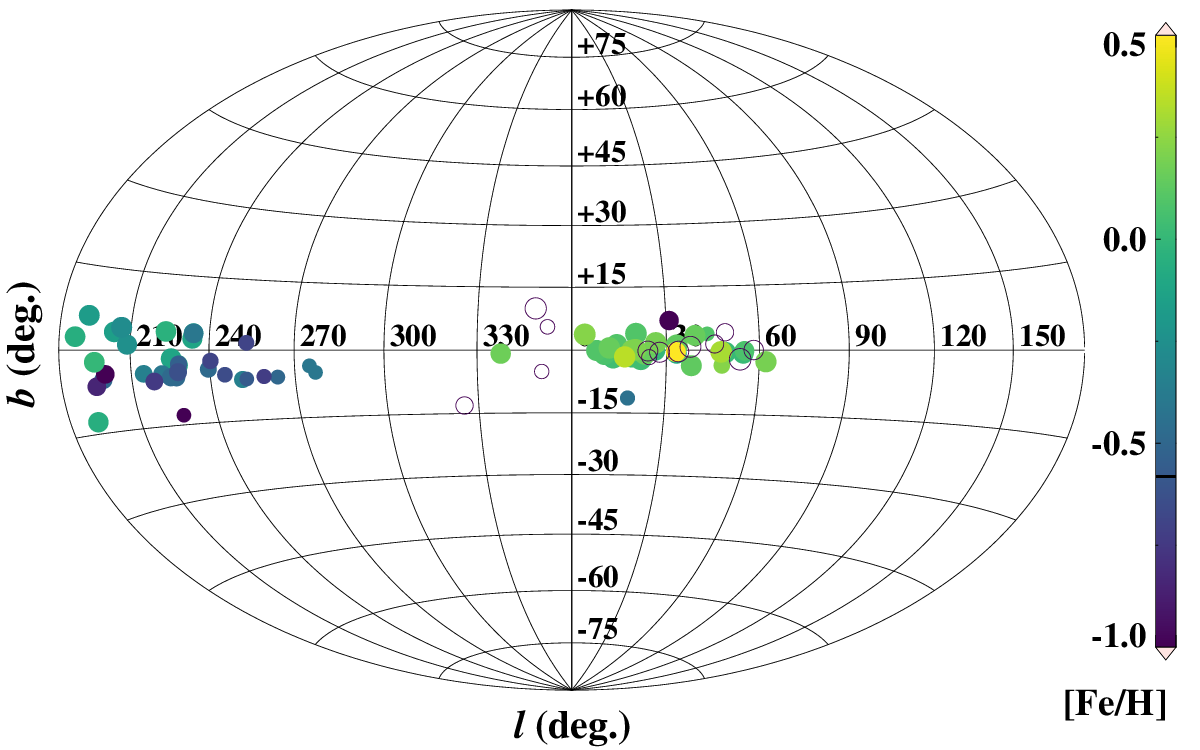}
        \caption{Spatial distribution of classical Cepheids used in this work in Galactic coordinates. The smaller symbol sizes represent distant Cepheids. Open circles represent stars with no metallicity measurements. The colour bar represents metallicity.}
  \label{fig:gal_coord}
\end{figure}

The sample of stars was selected to obtain complementary time-series data for Cepheids that have high-resolution spectroscopic observations as part of the C-MetaLL survey \citep{ripepi2021}.
Our initial sample consists of 80 Cepheids including 49 fundamental mode (FU) and 31 first-overtone mode (FO) stars. 
The majority of our Cepheids were identified as variables in the {\it Gaia} data releases. However, some of our 
initial targets, which were included in \citet{ripepi2021}, come from the ASAS-SN or the OGLE survey \citep{udalski2018, jayasinghe2018}. The pulsation periods, epochs of maximum brightness, 
intensity-weighted mean magnitudes in $G$, $G_{BP}$, and $G_{RP}$ bands, and $G$ band pulsation amplitudes for 73 Cepheids were taken from {\it Gaia} DR3 Cepheid sample \citep{ripepi2022}. For remaining seven Cepheids, only photometric mean magnitudes in {\it Gaia} filters were taken from the source catalogue \citep{vallenari2022} since they were not included as variables in {\it Gaia}.
The periods and epochs of maximum brightness, and the $V/I$ band amplitudes for these stars were taken from the OGLE (OGLE-GD-CEP-0307/1210/1227/1286) or ASAS-SN 
(J163656.55-322102.2, J184816.35+004903.4 / V0912 Aql, J193837.90+172322.8) catalogues \citep{udalski2018, jayasinghe2018}. Figure~\ref{fig:gal_coord} displays the location of Cepheid variables used in this work. 

\begin{figure}
  \includegraphics[width=0.96\columnwidth]{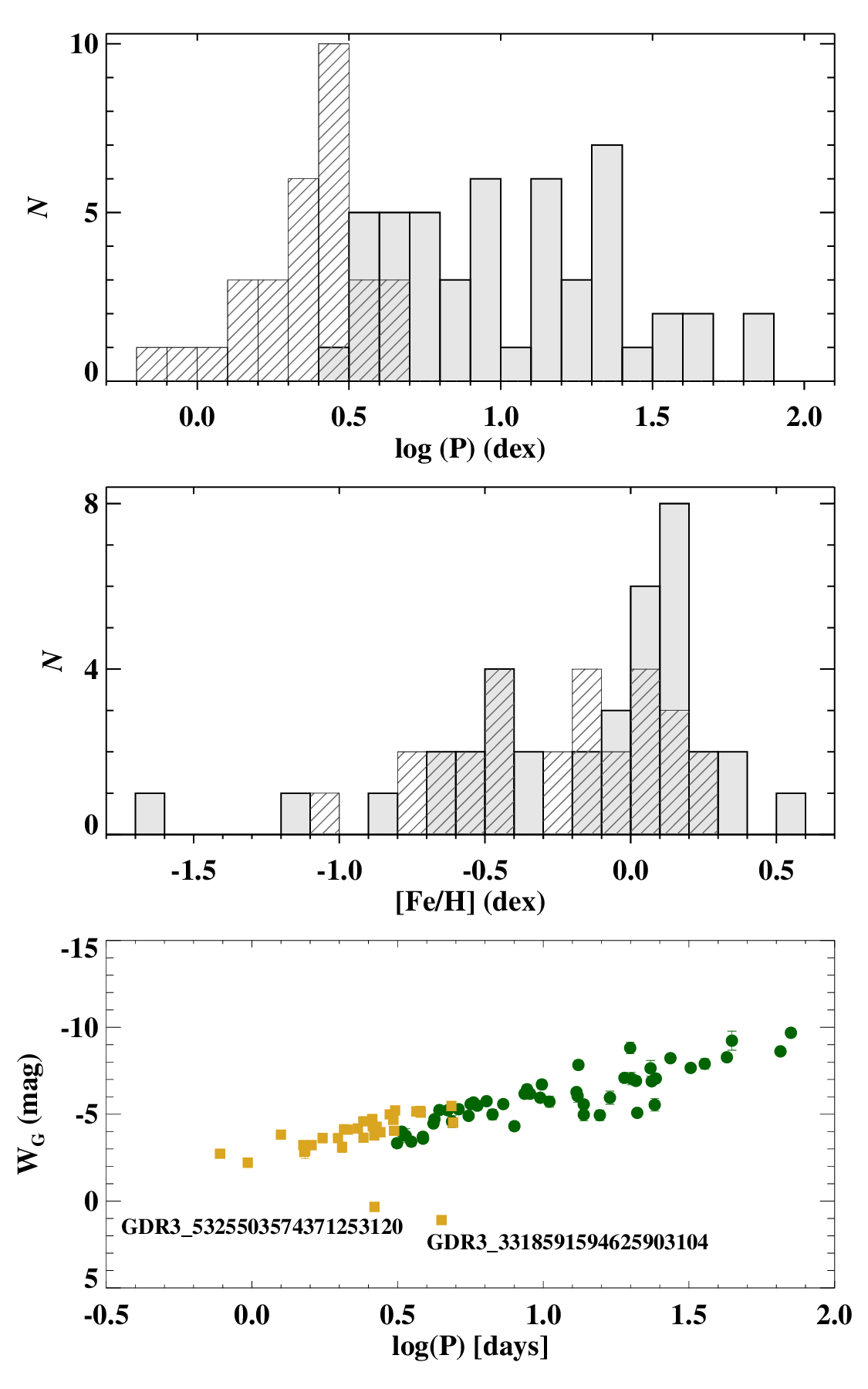}
        \caption{The period and metallicity distributions, and PW relation for MW Cepheids. {\it Top:} Period distribution of 80 candidate Cepheids in the sample studied in this work. 
	The filled and lined histograms correspond respectively to FU and FO Cepheids. {\it Middle:} Histogram of metallicities for 65 Cepheids, of which 59 have homogeneous high-resolution spectroscopic metallicities from the C-MetaLL survey \citep{ripepi2021,trentin2023}. {\it Bottom:} Period-Wesenheit relation in {\it Gaia} filters for the sample of 80 stars. The circles and squares represent fundamental and first-overtone mode Cepheids, respectively. Two extreme outliers are marked with their {\it Gaia} DR3 source ID.}
  \label{fig:hist_all}
\end{figure}

Figure~\ref{fig:hist_all} displays the histogram of period distribution of Cepheids in our sample. The FU mode Cepheids cover a period range of 3.15 to 70.80 days while FO Cepheids have periods 
between 0.78 and 4.91 days. These Cepheids are bright and cover an apparent magnitude range between $G=8.89$ and $G=14.78$ mag. The distances to our Cepheids range from 0.5 kpc to 19.7 kpc 
with a median distance of 3.6 kpc. High-resolution spectroscopic metallicities for 59 Cepheids were already obtained as part of the C-MetaLL survey \citep{ripepi2021, trentin2023}. Among the remaining 
stars, six more Cepheids have medium-resolution spectroscopic metallicities from {\it Gaia} Radial Velocity Spectometer  \citep[RVS,][]{blanco2022}. However, four of these six Cepheids have
poor RVS quality flags and were assigned a large uncertainty of 0.5~dex \citep{trentin2023a}. The middle panel of Figure~\ref{fig:hist_all} shows the histograms of available metallicities for all 65 Cepheids.

To confirm the identification and classification of 80 Cepheids in our sample, we derived their PW relation, $W_G=1.90(G_{BP}-G_{RP})$, in the {\it Gaia} bands \citep{ripepi2019}. The distances to these Cepheids from \citet{bailer2021} were used to obtain absolute Wesenheit magnitudes. We noticed two outlier stars (5325503574371253120/OGLE-GD-CEP-0307, 3318591594625903104/OGLE-GD-CEP-1286 in the PW relation shown in the bottom panel of Figure~\ref{fig:hist_all}. OGLE-GD-CEP-1286 is classified as rotating variable in the ASAS-SN survey (J061105.05+045303.5).  The parallax uncertainties of these stars are less than $3\%$ and these were not classified as Cepheids in the {\it Gaia} data. Therefore, we excluded these two stars from our final sample, which consists of 78 stars with 49 FU and 29 FO Cepheid variables.

\subsection{Observations and data reduction}

Our multiband observations of Cepheid variables were obtained between February 2021 and September 2022 using the Rapid Eye Mount (REM{\footnote{\url{http://www.rem.inaf.it/}}}) Telescope  located in La Silla, Chile. REM is a 60~cm diameter fast reacting telescope with two parallel imaging instruments: ROS2,  a visible imager with four simultaneous passbands ($griz$), and REMIR,  an infrared ($JHK_s$) imaging camera. The pixel scale of ROS2 and REMIR is 0.58 and 1.2 arcsec/pixel, respectively, for a total field of view of about $10\times$10 arcmin. The time-series observations were obtained in a monitoring mode where each Cepheid was observed once per night, when visible during the semester. Our observations for each epoch consisted of three dithered frames (exposures) in optical and a minimum of five dithered frames in NIR. 
The exposure times in optical bands were optimized depending on the brightness of the sources, and range from two to 240s. The individual exposures were relatively short in NIR (between one and ten seconds), such that the background sky variations were negligible. For fainter targets, multiple sequences of five dithered exposures were taken in NIR bands. These Cepheids were observed on average on 24 different nights with the number of observations varying between ten and 60. 

The optical and NIR science images and associated calibrations (bias and flat frames) were downloaded from the REM data Archive{\footnote{\url{http://ross.oas.inaf.it/REMDB/}}}. For optical images, the bias subtraction and flat fielding were performed using monthly calibration frames. For NIR images, pre-processed (dark subtraction and flat fielding) images were downloaded including the sky background and co-added images. However, the co-added NIR images were not used because several bright residual patterns were present which likely resulted from over-subtraction of sky background in several pixels. These residual variations were normalized using \texttt{SExtractor} \citep{bertin2006} in the sky background images before applying background subtraction. Therefore, we used individual dithered frames in both optical and NIR bands for photometric data reduction.

\begin{figure}
  \includegraphics[width=0.98\columnwidth]{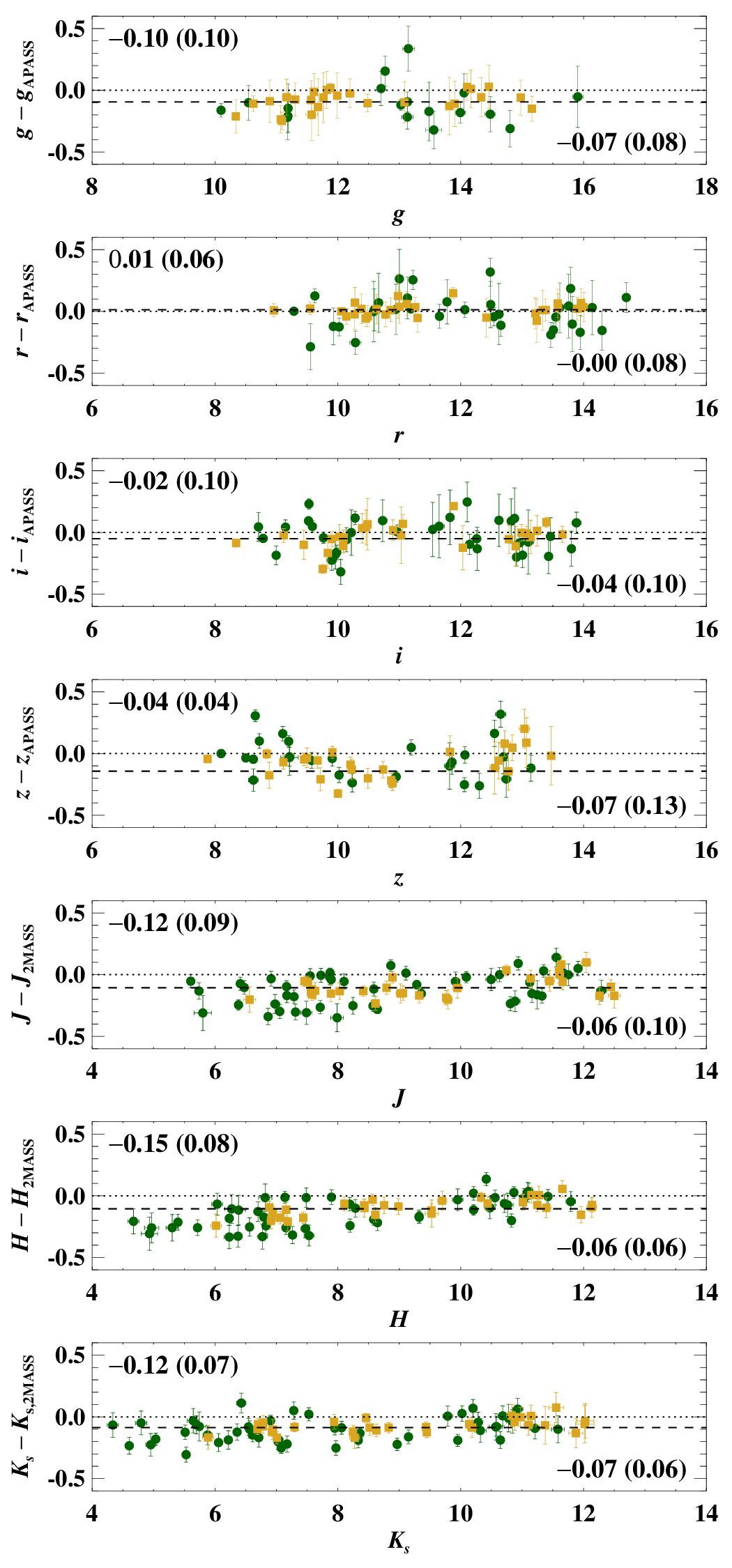}
  \caption{Comparison of the calibrated mean magnitudes of Cepheids in this work with the magnitudes from APASS DR10 (in $griz$) and 2MASS (in $JHK_s$). The dashed and dotted lines respectively  represent zero and mean offset. The mean (and the standard deviation) offsets for the bright ($K_s < 8.5$~mag) and faint ($K_s \ge 8.5$~ mag) Cepheids are also mentioned at the top left and bottom right of each panel.}
  \label{fig:delmag}
\end{figure}

\subsection{Photometry}

The photometry was performed on all dithered images separately in each filter using the \texttt{DAOPHOT/ALLSTAR} \citep{stetson1987} and \texttt{ALLFRAME} \citep{stetson1994} routines. In a given filter, all point-like sources were first identified using \texttt{SExtractor} to determine a full width at half-maximum (FWHM). An aperture photometry was then performed using  \texttt{DAOPHOT} with an aperture equal to the median FWHM. Up to ten brightest, non-saturated stars were used to construct an empirical point-spread-function (PSF) excluding sources closer to the detector edges. We selected the first-frame (images of the first-night observations) as reference frame to create a star-list. In cases where there were not enough detections ($<10$) in the first-frame, we visually selected images taken on another night to create a reference star list. The PSF photometry was obtained using  \texttt{ALLSTAR} on all images and frame-to-frame coordinate transformations between the reference image and all epoch images were derived using \texttt{DAOMATCH/DAOMASTER}. This procedure required the most intervention because of the lack of a sufficient number of stars in images taken in poorer weather conditions resulted in inaccurate transformations. Those frames were excluded for which no transformation could be obtained due to poorer statistics or lack of significant overlap with the reference frame. Finally, the reference-star list and the derived coordinate transformations were used as input to \texttt{ALLFRAME} for performing PSF fitting across all the frames, simultaneously. The photometry on different nights was internally calibrated to the reference frame using secondary standards that have small photometric uncertainties and no epoch-to-epoch variability. These thresholds on photometric uncertainties and epoch-to-epoch variability varied between 0.01 and 0.1~mag depending on the number of sources within the field of view and their photometric precision.

The photometric calibration was performed using {\it Gaia} synthetic photometry \citep[GSP, ][]{montegriffo2023} in Sloan filters ($griz$) and 2MASS \citep[][]{skrutskie2006} catalogue in the $JHK_s$ bands. We selected homogeneous GSP catalogue for calibrating optical photometry because there were not enough common stars for several sources in the American Association of Variable Star Observers (AAVSO) Photometric All-Sky Survey (APASS) DR10 \citep{henden2019}. Moreover, most of these bright Cepheids were saturated in the Sloan Digital Sky Survey catalogue. For each Cepheid location, all sources from GSP and 2MASS catalogues were extracted within the field of view of $10\times10$ arcmin. Our photometric catalogues in a given filter were then cross-matched with GSP or 2MASS within an initial tolerance of $1.0\arcsec$. The common sources were used to determine a zero-point offset to calibrate our instrumental magnitudes to Sloan or 2MASS filters. The number of calibrating stars varied between five and 117 for different star or filter, and there were fewer common stars in $gz$-bands. The lack of number of stars covering a wide colour range precluded us from solving for a colour-term in the calibration equation. The uncertainties in the magnitude zero-point offset were propagated to the instrumental magnitudes for each Cepheid variable. 

Figure~\ref{fig:delmag} shows a comparison of our calibrated mean magnitudes of Cepheids (see Section~\ref{subsec:meanmag}) with the magnitudes from APASS DR10 and 2MASS for $griz$ and $JHK_s$ filters, respectively. The offsets are small in $ri$ bands and increase in $gz$ bands and at NIR wavelengths, where magnitudes from the literature are based on random-epoch observations. In the case of NIR magnitudes, the offsets increase significantly at the bright end ($K_s < 8.5$ mag). We note that the stars with $4 < K_s < 8.5$ mag saturate in the primary $1.3$s 2MASS exposure and their NIR magnitudes come from aperture photometry on 51ms exposure images \citep{skrutskie2006}. The largest offset occurs at the brightest end in the $H$-band. For $H < 6.5$ mag, the median errors in the REMIR $J/H/K_s$ magnitudes of these Cepheids are 0.05/0.09/0.08 mag, hinting at possible non-linearity and saturation at the bright end in $HK_s$. The typical uncertainties on 2MASS magnitudes are 0.02-0.03 mag, and thus the offsets at the bright end are still within $2\sigma$ of the combined REMIR and 2MASS uncertainties. However, we also noted increased scatter in the light curves around maximum light for Cepheids with $K_s < 5$~mag. In these cases, the brightest source in the REMIR images is typically the target Cepheid, and the PSF is generally constructed from fainter stars. The aperture photometry may be a better option for these brightest Cepheids, and will be explored in the subsequent photometric studies of the C-MetaLL survey.

\begin{figure*}
\centering
  \includegraphics[width=0.95\textwidth]{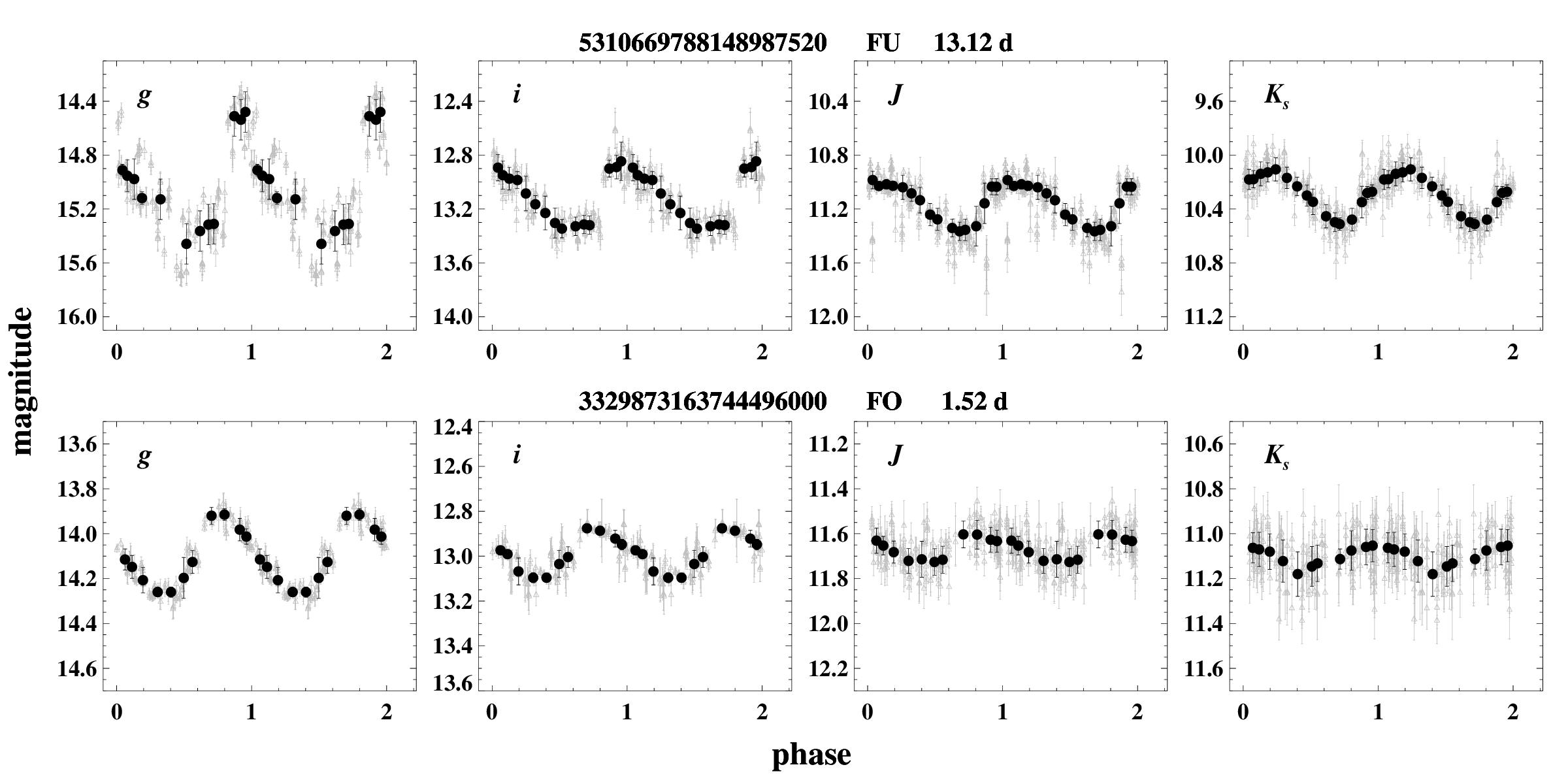}
        \caption{Example light curves of FU (top) and FO mode (bottom) Cepheids based on all the photometric data (in grey) from the dithered frame images. The light curves are shown in 
        two optical ($gi$) and two NIR ($JK_s$) filters. The black circles represent the weighted mean values in a given phase bin and the errors represent their standard deviations. The range of $y$-axis 
        is the same in each band for a given Cepheid. The {\it Gaia} source ID, pulsation mode, and the periods are listed at the top of each panel.}
  \label{fig:bin_lcs}
\end{figure*}

\begin{figure*}
  \centering
  \begin{tabular}{ccc}
    \includegraphics[width=0.3\textwidth]{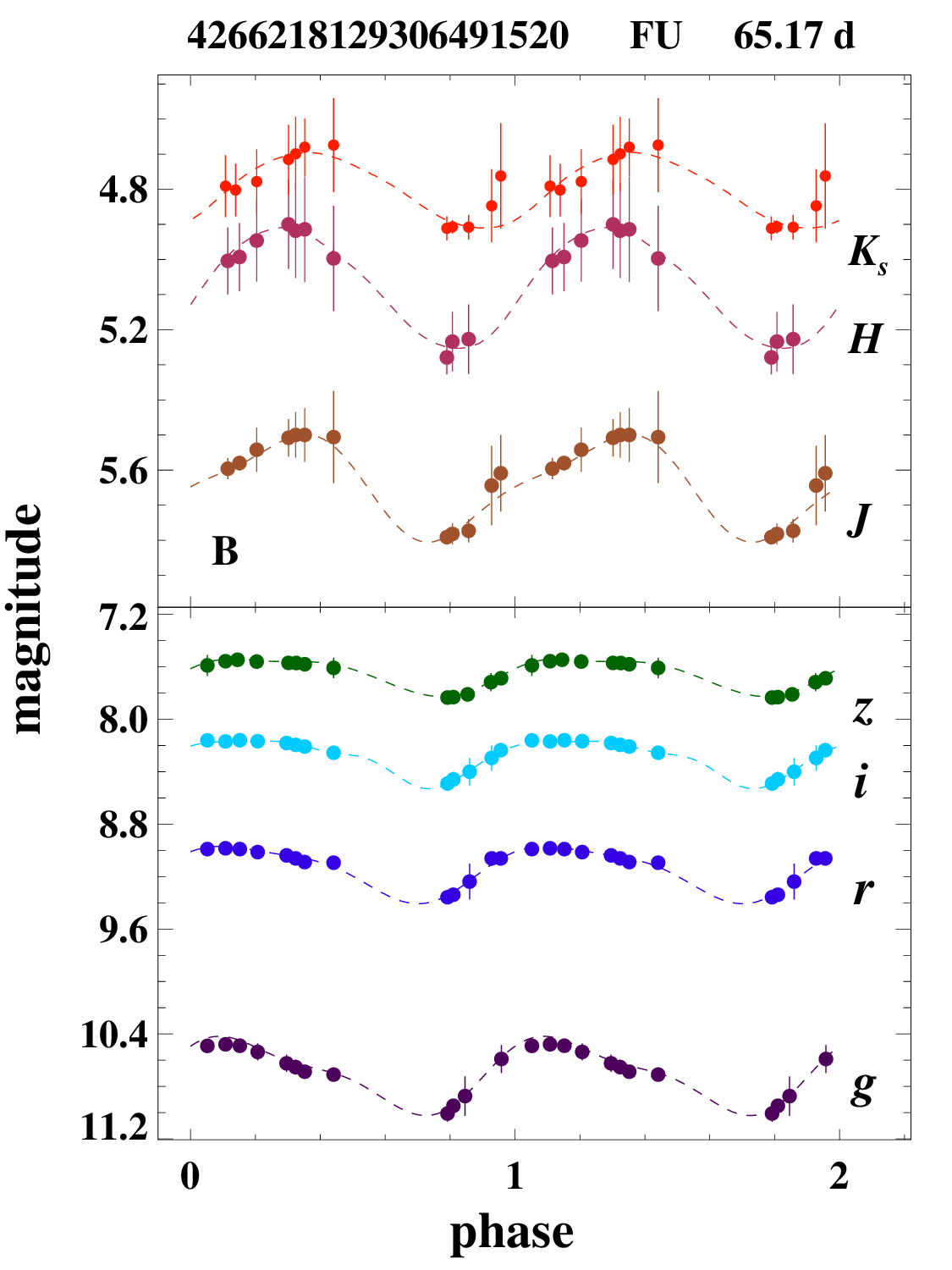} &    
    \includegraphics[width=0.3\textwidth]{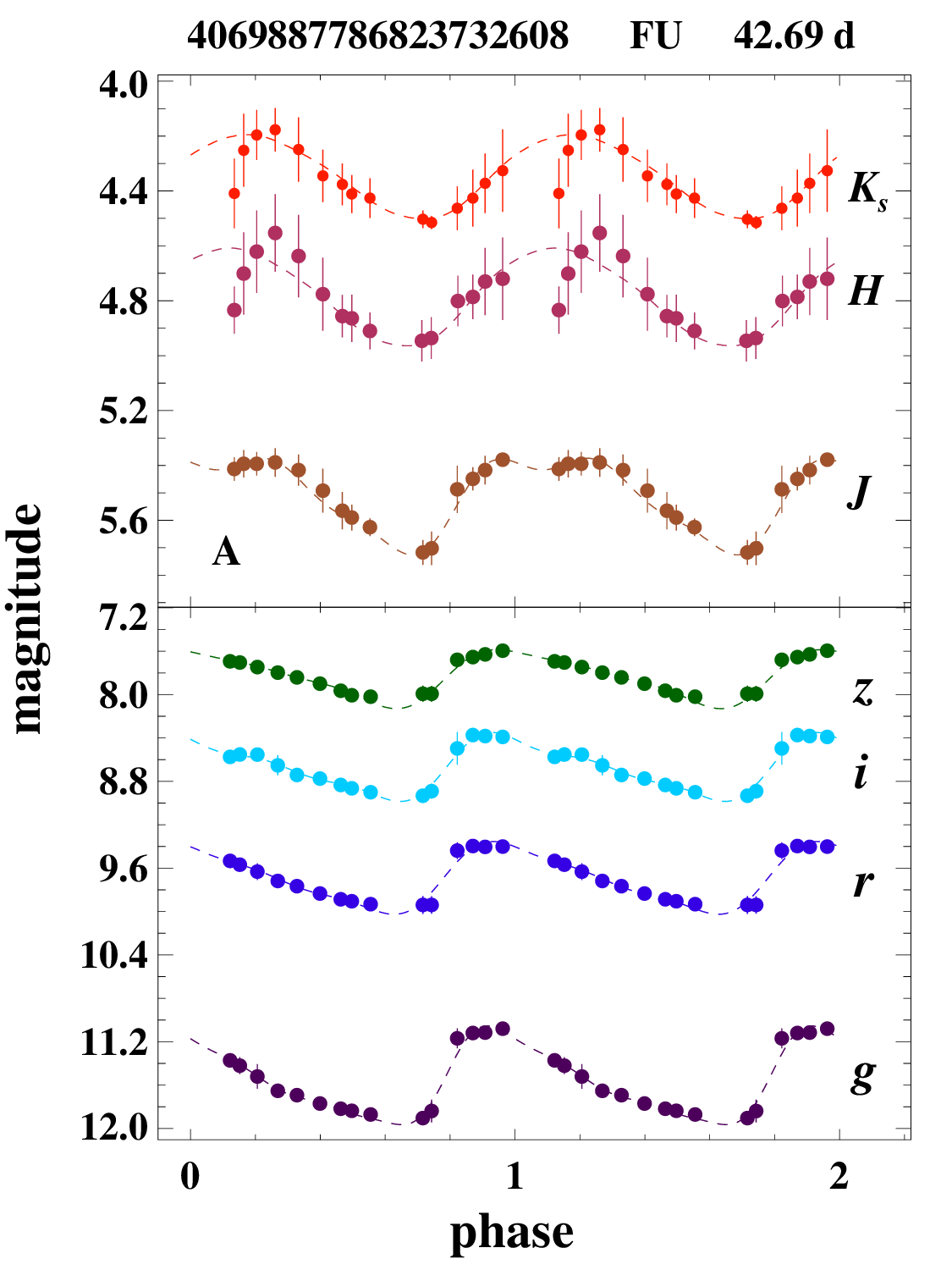} &
    \includegraphics[width=0.3\textwidth]{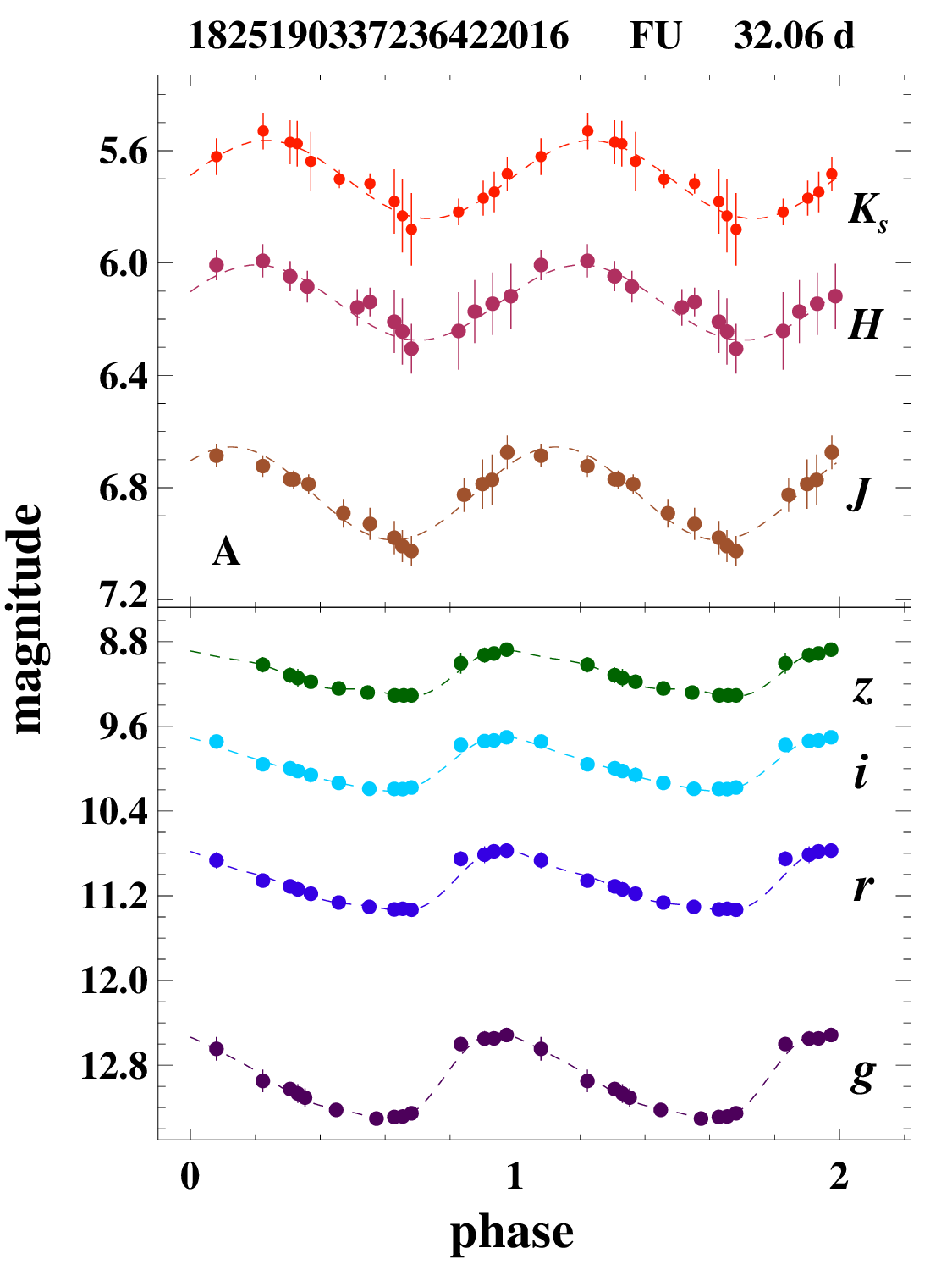}\\
          \includegraphics[width=0.3\textwidth]{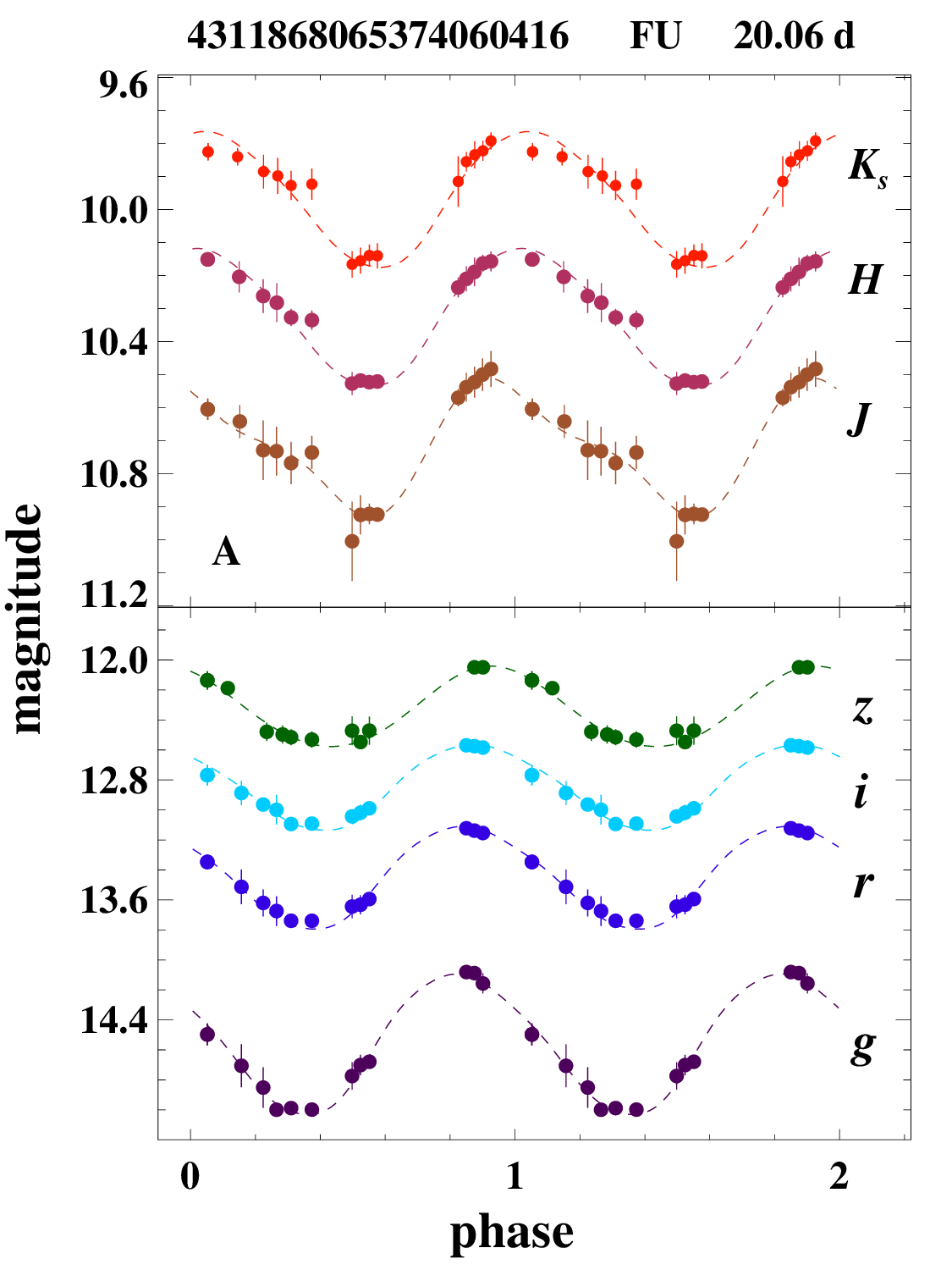} &
  \includegraphics[width=0.3\textwidth]{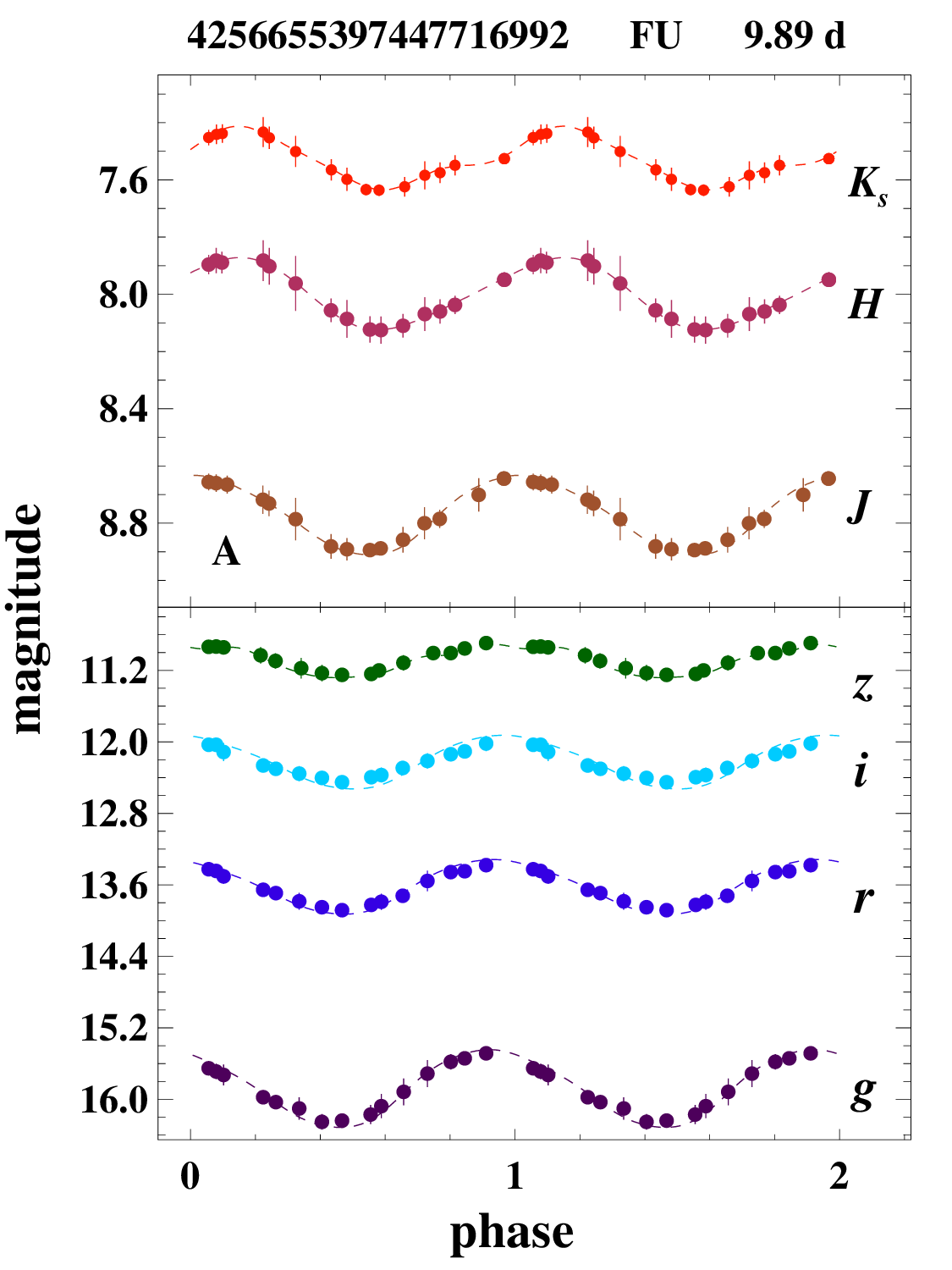} &
    \includegraphics[width=0.3\textwidth]{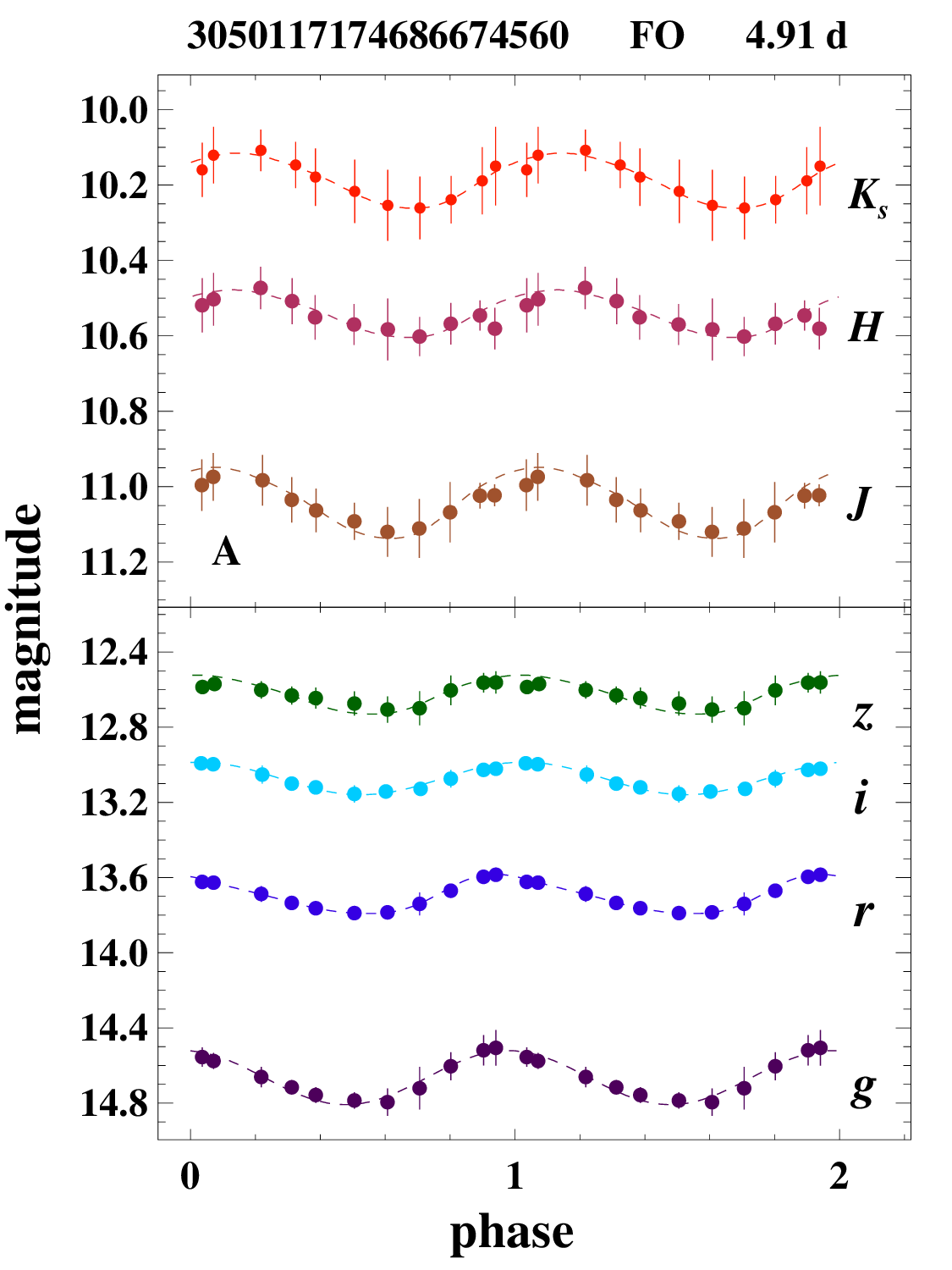} \\
          \includegraphics[width=0.3\textwidth]{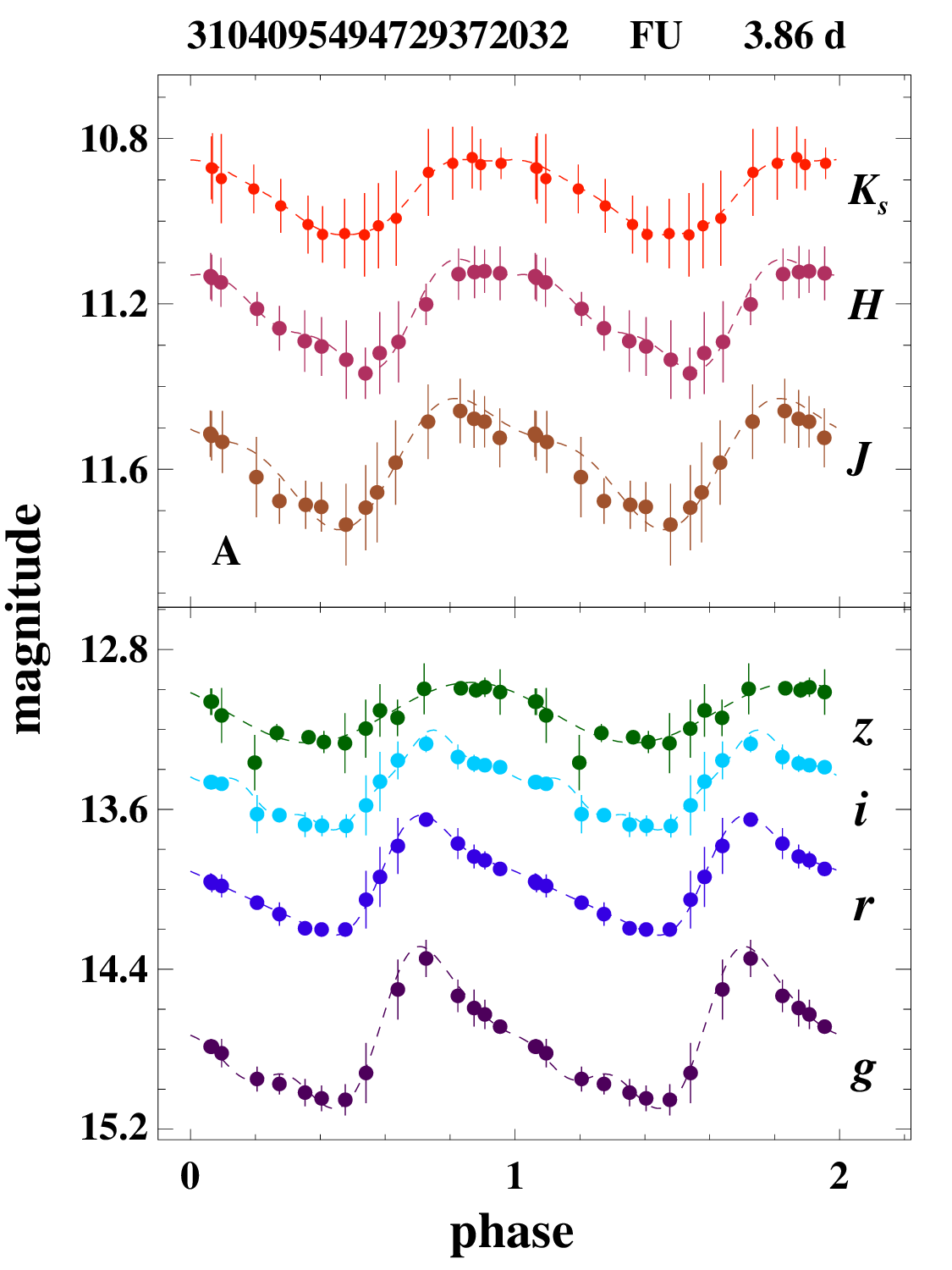} &
  \includegraphics[width=0.3\textwidth]{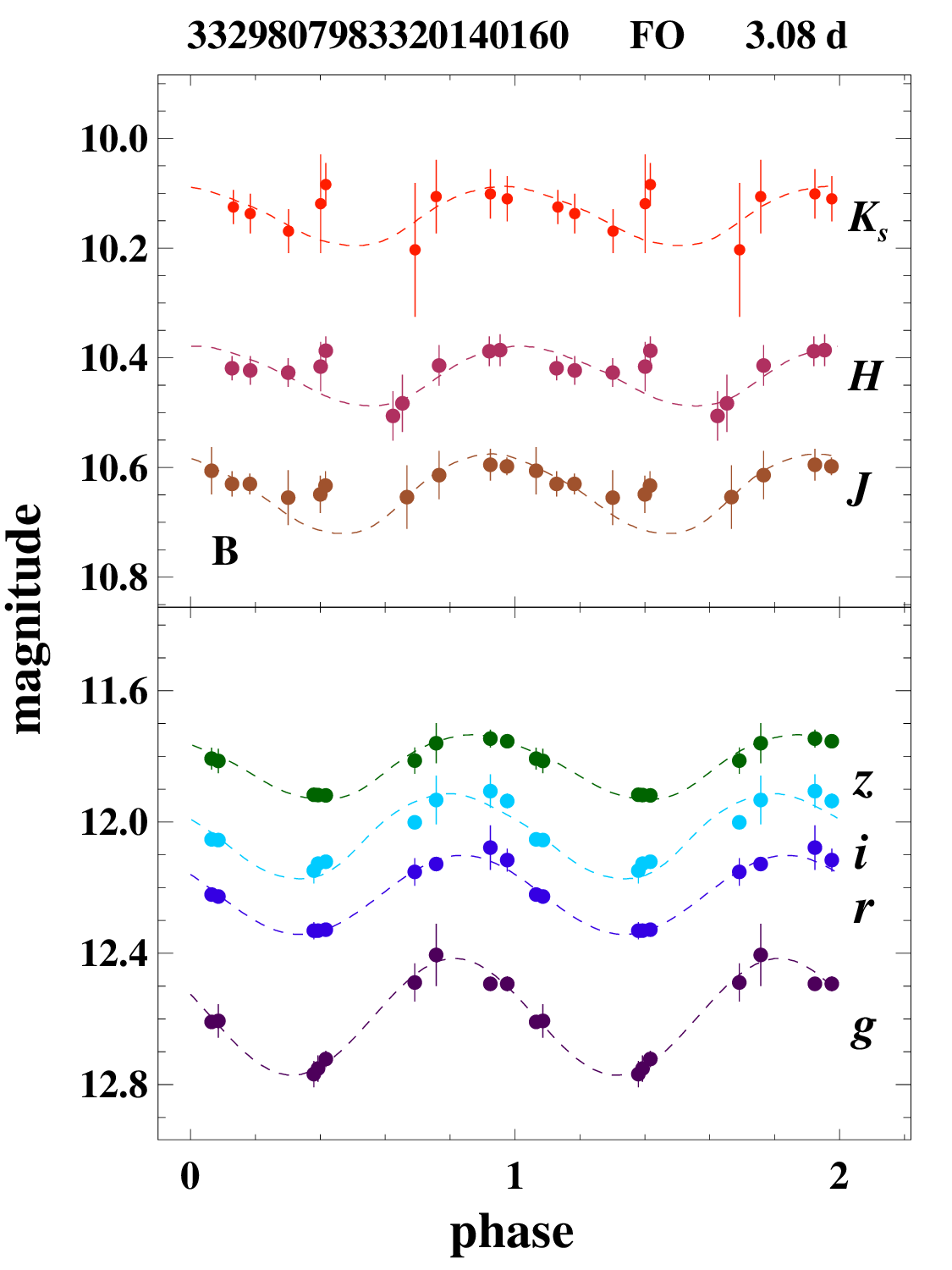} &
    \includegraphics[width=0.3\textwidth]{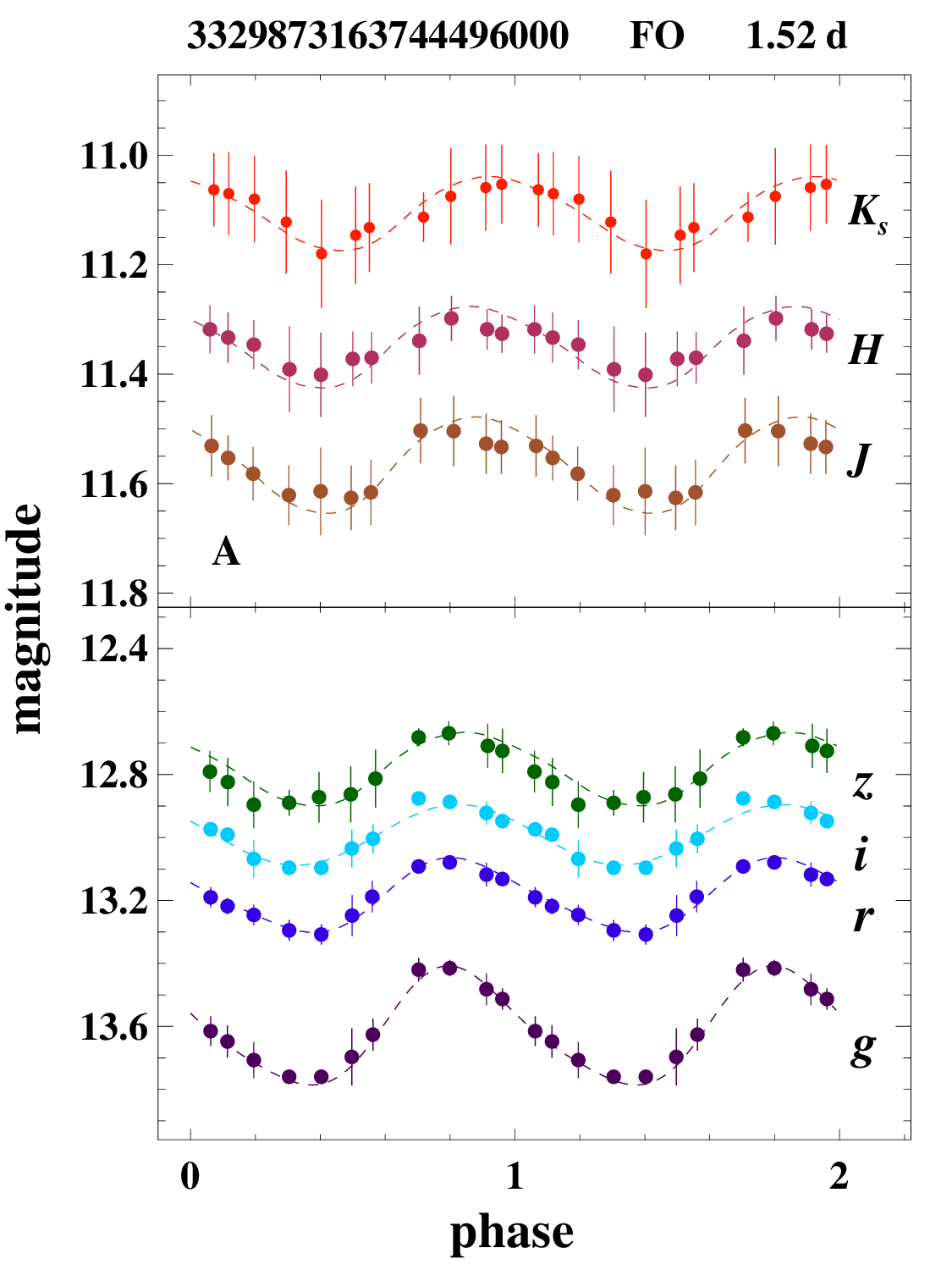} \\
\end{tabular}
        \caption{Representative phase-binned light curves of FU and FO Cepheids with varying periods at all wavelengths ($grizJHK_s$). The magnitudes in $g, r, J, H$ bands were offset by 
        $-0.5, -0.2, -0.1, +0.1$ for visualization purposes. The best-fitting Fourier series fits are also shown as dashed lines. The {\it Gaia} Source ID, pulsation mode, and period are listed at the top of each panel. The light curve quality flag is given at the bottom left of each panel. The uncertainties in magnitudes at a given phase point also include
        the scatter at that phase in the original light curves.}
             \label{fig:cep_lcs}
\end{figure*}

\label{app:table}
\begin{table*}
\begin{center}
\caption{Pulsation properties and multiband photometric intensity-averaged magnitudes of Galactic Cepheid variables. \label{tbl:phot_data}}
\small
\begin{tabular}{cccrrrrrrrrrrr}
\hline\hline
        {ID}  & {$P$} & {M}  & {\bf $g$} & {\bf $r$}  & {\bf $i$} & {\bf $z$} & {\bf $J$}  & {\bf $H$} & {\bf $K_s$} & QF & {\bf $E_{BV}$} & {[Fe/H]} & Ref.$^b$\\ 
        &   days   &         & \multicolumn{7}{c}{mag}                                                              &     & mag     & dex & \\
\hline
1825190337236422016$^{~}$&   32.06123&   FU&    13.41&    11.27&     9.96&     9.10&     6.91&     6.04&     5.70& A&     1.54&$     0.03$& R21\\
                             &         &   &    0.06&     0.05&     0.04&     0.06&     0.06&     0.09&     0.07& &     0.07&$     0.11$& \\
1825428480248517376$^{~}$&    5.93870&   FU&    14.05&    12.49&    11.54&    10.95&     9.29&     8.58&     8.36& A&     1.20&$     0.08$& RVS\\
                             &         &   &    0.03&     0.03&     0.05&     0.03&     0.03&     0.04&     0.03& &     0.07&$     0.50$& \\
1834213582473654144$^{~}$&    5.63099&   FU&    13.21&    10.95&     9.90&     9.11&     7.72&     7.14&     6.91& B&     1.20&$     0.20$& R21\\
                             &         &   &    0.06&     0.03&     0.08&     0.04&     0.05&     0.04&     0.03& &     0.07&$     0.15$& \\
2018385387883485056$^{~}$&   70.79719&   FU&    12.58&    10.29&     8.94&     8.11&     5.80&     4.93&     4.60& B&     1.61&    ---& ---\\
                             &         &   &    0.04&     0.09&     0.05&     0.06&     0.14&     0.13&     0.06& &     0.07&    ---& \\
2936063665309240576$^{~}$&    1.97524&   FO&    14.33&    13.59&    13.12&    12.84&    11.61&    11.15&    10.97& A&     0.54&$    -0.55$& T22\\
                             &         &   &    0.03&     0.04&     0.04&     0.06&     0.07&     0.06&     0.08& &     0.07&$     0.17$& \\
2936165984303583360$^{~}$&    3.52263&   FU&    14.27&    13.47&    13.01&    12.74&    11.56&    11.03&    10.84& A&     0.54&$    -0.65$& T22\\
                             &         &   &    0.05&     0.05&     0.05&     0.08&     0.07&     0.07&     0.09& &     0.07&$     0.13$& \\
2936274153063501824$^{~}$&    6.38872&   FU&    10.10&     9.29&     8.79&     8.50&     7.17&     6.70&     6.57& B&     0.52&$    -0.11$& R21\\
                             &         &   &    0.03&     0.02&     0.03&     0.03&     0.04&     0.07&     0.05& &     0.07&$     0.11$& \\
2940212053953709312$^{~}$&   16.93958&   FU&    13.33&    12.62&    12.27&    12.06&    10.88&    10.48&    10.32& B&     0.18&$    -1.10$& T22\\
                             &         &   &    0.04&     0.03&     0.06&     0.06&     0.08&     0.09&     0.09& &     0.08&$     0.19$& \\
2947876298535964416$^{~}$&    2.41590&   FO&    13.90&    13.22&    12.79&    12.55&    11.44&    11.01&    10.88& A&     0.42&$    -0.48$& T22\\
                             &         &   &    0.05&     0.05&     0.05&     0.08&     0.04&     0.05&     0.06& &     0.07&$     0.11$& \\
4278570592634887552$^{a}$&    4.40050&   FU&    ---&    10.58&     9.37&     8.63&     6.98&     6.23&     6.06& B&     1.48&$     0.14$& R21\\
                             &         &   &   ---&     0.06&     0.05&     0.13&     0.07&     0.09&     0.06& &     0.10&$     0.11$& \\
\hline
\end{tabular}
\end{center}
        \footnotesize{{\bf Notes:} The {\it Gaia} DR3 Source IDs, periods, and pulsation modes (M) (fundamental, FU,  or first-overtone, FO) are given in the first three columns.
        QF is the  quality flag of the light curve (A, B, C for best, good, poor). $E_{BV}=E(B-V)$.\\
        $^a$The seven Cepheids that were not classified as variables in the {\it Gaia} DR3 \citep{ripepi2022}; see Section \ref{sec:data_sample} for details.\\
        $^b$References for metallicities: R21  \citep{ripepi2021}, T22  \citep{trentin2023}, RVS  \citep{blanco2022}.\\
 This table is available in its entirety in machine-readable form.
        }
\end{table*}

\section{Optical and near-IR light curves of Cepheid variables}
\label{sec:var}

\subsection{Pulsation periods and the phase-binned light curves}

The photometric light curves of Cepheids in both optical ($griz$) and NIR ($JHK_s$) filters were utilized to determine their pulsation periods using the multiband periodogram of \citet{saha2017}.
This hybrid algorithm for period determination using sparsely sampled light curve data at multiple wavelengths was employed to search periods in the range from 0.1 to 100 days with a stepsize of 0.001 days. An average difference of 0.002 days with a scatter of 0.012 days was found between our periods and those adopted from the literature after excluding a few spuriously determined periods. The difference in periods exceeded 0.1 days for 17 Cepheids, all of which either had large phase gaps in their light curves or poor light curve quality in at least one filter. Since all the 
Cepheids in our sample are bright ($G<14.5$~mag), we adopted literature periods primarily from {\it Gaia} DR3 \citep{ripepi2022} that were based on much longer temporal baseline than our observations.

Multiwavelength light curves were phased using the adopted periods and the epochs of maximum brightness. Since we obtained multiple dithered frames in both optical and NIR filters and performed photometry on individual frames, there are often several data points at a given phase albeit with larger uncertainties due to lower signal-to-noise of individual exposures. Furthermore, the scatter at a given phase is larger for stars with a small number of secondary standards within their imaged field of view since these were used to obtain frame to frame transformations and also to calibrate
the photometry. Figure~\ref{fig:bin_lcs} displays example light curves of FU and FO Cepheids with all epoch observations (in grey). The scatter in the light curves is evident which comes from larger errors due to low signal-to-noise of dithered frames and uncertainties in relative zero-points of photometry from different nights. Therefore, we decided
to bin the light curves in phase improving the accuracy and precision of photometric data points in a given phase bin. We computed sliding mean values with a bin width of $0.1$ in phase with 15/10 steps for
FU/FO Cepheids. While 10 phase averaged data points were sufficient for sinusoidal nature of FO Cepheid light curves, we used 15 steps for FU Cepheids to fully recover the saw-tooth feature of their
light curves. We found these choices recovered the sharp extrema of the light curves and also yielded the least amount of scatter between consecutive points. A weighted mean of all magnitudes was 
obtained from dithered frames within a given phase bin and the standard deviation of the robust mean was propagated to the photometric uncertainties. 

Figure~\ref{fig:bin_lcs} also shows the phase-binned light curves of Cepheids overplotted on the original data points. While the variability was quite evident in the original light curves, it is more 
pronounced in the phase binned light curves. The phase-binning procedure was particularly useful for low-amplitude FO variables for which the larger uncertainties on photometry from individual dithered 
frames often dominated variability amplitudes in NIR bands. The light curves constructed from the phase binning procedure are used for further analysis in this work. These light curves were visually inspected
to assign a quality flag of A,B, and C to the best, good, and poor light curves depending on the scatter and the phase coverage of the light curves. There are 52/22 stars with best and good (`A and B') 
quality flags, and only four stars have poor light curves. We note that the entire sample of Cepheids was used in the subsequent analysis irrespective of their light curve quality flags.

\begin{figure*}
\centering
  \includegraphics[width=0.93\textwidth]{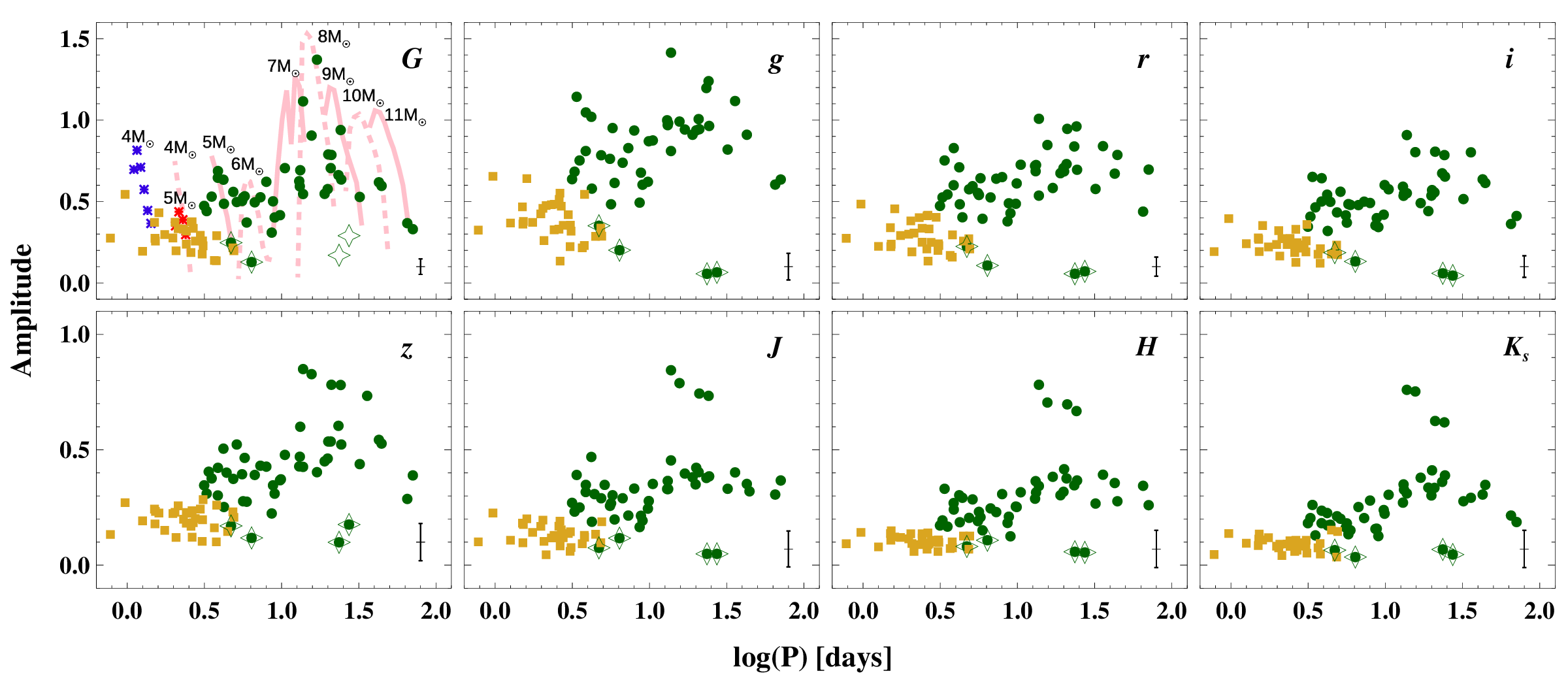}
        \caption{Period-amplitude diagrams for Cepheids in multiple bands. The filled circles and squares represent FU and FO Cepheids, respectively. In the case of {\it G} band, the theoretically predicted amplitudes for metallicities ($Z=0.02$) representative of Cepheids in the MW are also shown for different masses \citep{desomma2022}. The solid and dashed lines represent FU Cepheid models, while asterisks  show FO models. In $grizJHK_s$ bands, overplotted Cepheids in open star symbols are discussed in  Section \ref{sec:pamp_plot}. } 
  \label{fig:ampl}
\end{figure*}

\subsection{Light curve fitting and mean magnitudes}
\label{subsec:meanmag}
The light curve templates for Cepheids are not available in optical Sloan ($griz$) filters, but are available in NIR bands \citep[e.g.][]{inno2015}. Since our light curves are phase-binned, we do not
expect to recover small features such as bump near the maximum light, particularly at NIR wavelengths where light curves are more sinusoidal. Therefore, we decided to analyse multiband light curves with
the Fourier decomposition method \citep[e.g.][]{bhardwaj2015, bhardwaj2017} and not rely on a particular set of templates. The phase-binned light curves were fitted with a Fourier sine series fit in 
the following form:
$$ m_\lambda = m_{0,\lambda} + \sum_{k=1}^{N} A_{k,\lambda} \sin(2k\pi\cdot\Phi_\lambda + \phi_{k,\lambda}), $$

\noindent where $m_\lambda$ is the magnitude as a function of phase ($\phi_\lambda$) at a given wavelength ($\lambda$). The mean magnitudes ($m_{0}$) and Fourier coefficients ($A_k,~\phi_k$) were determined
using a Fourier sine series with varying order of fit ($k=1$ to 5). A minimum order of $k=1$ was chosen to obtain purely sinusoidal fits, which are ideal for FO Cepheids, particularly at NIR wavelengths. 
Since the phase-binned light curves do not necessarily provide enough data points around their extrema, we find that a fifth order Fourier series was sufficient to fit the main saw-tooth 
characteristics and Hertzsprung progression \citep{hertzsprung1926} features. Furthermore, limiting higher order fit to $k=5$ also avoided possible numerical ringing in case of larger phase gaps. 
The best order of fit was determined using the Baart's criteria \citep{baart1982}.

Figure~\ref{fig:cep_lcs} displays the Fourier-fitted light curves in $grizJHK_s$ filters for FU and FO Cepheids with varying periods. The saw-tooth light curve shape for a short-period 
($P=3.86$~days) FU Cepheid and flatter maxima for Cepheids with periods close to 10 days can be noted despite the phase-binning. Most Cepheids with periods longer than 20 days typically have similar
near-sinusoidal light curve shapes with decreasing amplitude as a function of period. In the case of FO Cepheids, sinusoidal Fourier series with $k=1$ fitted well the majority of stars at all wavelengths.
The best-fitting Fourier fits were used to derive intensity-averaged magnitude and peak-to-peak amplitude for all Cepheids except for four stars in $g$ band and one star in $K_s$ band. These Cepheids with lack of measurements in one filter did not have enough light curve phase points for a Fourier-fit or mean-magnitude determinations. The photometric mean magnitudes of all stars are
listed in Table~\ref{tbl:phot_data}. 

\subsection{Period-amplitude diagrams}
\label{sec:pamp_plot}

Figure~\ref{fig:ampl} displays period-amplitude diagrams for Cepheids at multiple wavelengths. For reference, the {\it Gaia} $G$ band amplitudes, if available, are shown in the first panel. The predicted $G$ band amplitudes for Cepheid models representative of MW variables with $Z=0.02,~Y=0.28$ are also shown for different masses ($M=4-11M_\sun$). These models were computed adopting canonical mass-luminosity relation and a fixed mixing length parameter ($\alpha=1.5$) for both FU and FO mode Cepheids \citep[see][for details]{desomma2022}. In general, there is a good agreement between predicted and observed amplitudes for FU mode Cepheids with a maxima around $\log P \sim 1.3$~days. The short-period FO Cepheid models exhibit larger amplitudes than observations \citep{desomma2022}. We note that the predicted amplitudes are sensitive to the adopted composition, mass-luminosity relations, and the convective efficiency in the pulsation models. 

In $grizJHK_s$ period-amplitude diagrams, two low-amplitude FU Cepheids (open stars) with $\log P\sim 1.4$ days
can be seen as outliers in Figure~\ref{fig:ampl}. These variables were classified as FU Cepheids in the ASAS-SN survey, but their light curves show large scatter not typical of Cepheids. 
Given their small $V$-band amplitudes of less than 0.3 mag, no obvious periodic variations were seen in our photometry despite a good phase coverage. The amplitudes of phase-binned $grizJHK_s$ 
light curves are smaller than 0.1 mag, and these were not classified as variables in {\it Gaia} as well. We keep their classification as FU Cepheids since their mean magnitudes are well 
determined, and are used to probe their position on the PL relations. In NIR bands, four Cepheids with $1.1 < \log (P) < 1.4$ days exhibit relatively high amplitudes. These Cepheids have good-quality light curves and their independent $G$-band amplitudes also range between 0.78 and 1.12 mag.

In general, the decrease in amplitudes from the shortest-period to 10 days, an increase
in amplitude with periods up to 20 days, and a reverse trend for the longer period Cepheids is also seen, which is typical for FU Cepheids \citep{bhardwaj2015}.
The separation of FU and FO Cepheids on the period-amplitude plane is noted for all wavelengths, but is most distinct in $K_s$-band. However, two FU Cepheids (open stars) with $P < 10$ days and $K_s$-band
amplitudes of less than 0.1 mag are also populating FO amplitude cluster in all other bands. These Cepheids (5718760258978008320, 2936274153063501824) were classified by {\it Gaia} as FU mode
and have periods of 4.70 and 6.39 days, and $G$-band amplitudes of 0.35 and 0.20 mag, respectively.

\section{Absolute calibration of the Leavitt law} 
\label{sec:plrs}

\subsection{Gaia parallaxes}

The {\it Gaia} astrometric parallaxes were obtained for all Cepheids from \citet{vallenari2022} to calibrate multiband PL and PW relations. A few of these Cepheids are in the Galactic
disk and anti-centre direction, and therefore the parallaxes of these distant targets have large uncertainties. The median error on parallaxes is $9.4\%$, but 15 Cepheids have parallax
uncertainties of $>25\%$. Two Cepheids in our sample have negative parallaxes, and therefore the subsequent analysis is carried out in the parallax space. Several studies have found 
that the {\it Gaia} parallaxes are systematically smaller resulting in larger distances \citep{arenou2018, groenewegen2018, riess2021, bhardwaj2021, molinaro2023}. \citet{lindegren2021} provided 
a parallax zero-point correction recipe that derives a shift for an individual star based on its magnitude, colour, and ecliptic latitude. The parallax correction varies between $-0.075$ 
and $0.023$ mas for Cepheids in our sample. However, these parallax corrections are now known to over-correct parallaxes resulting in smaller distances \citep{riess2021, reyes2023, molinaro2023}. 
We adopted {\it Gaia} parallaxes after applying the suggested corrections by \citet{lindegren2021} and include an over-correction offset of $0.014$ mas from \citet{riess2021}. The 
parallax uncertainties were increased by $10\%$ following \citet{riess2021} and include parallax correction uncertainty of 0.005 mas added in quadrature \citep{lindegren2021}.

In addition to the uncertainties on parallaxes, we also looked at {\it Gaia} astrometric quality flags: renormalized unit weight error (RUWE) and goodness of fit (GOF). There are three stars
with RUWE $> 1.4$ and one of these also has a GOF $=13.98$ exceeding a threshold of 12.5 adopted by \citet{riess2021}. The largest RUWE among these three Cepheids is 1.55, and their parallax 
uncertainty are $<15\%$. Moreover, six Cepheids also fall near the sharp inflation point in the \citet{lindegren2021} formulae at G = 11 mag, and their parallax corrections may have larger
uncertainties. Given that our sample size is modest, we do not exclude these Cepheids from our analysis but treat them cautiously when deriving PL relations.

\subsection{Reddening and extinction corrections}

\citet{groenewegen2018} provided reddening values for more than 450 Cepheids that were primarily taken from \citet{fernie1995}. However, there are only two Cepheids (X Sct and V5567 Sgr) in our sample in common with \citet{groenewegen2018}. \citet{ripepi2021} derived new period-colour relations to estimate their intrinsic colours, and therefore reddening values. We used equation (4) of \citet{ripepi2021} to derive intrinsic $(V-I)_0$ colour for all Cepheids. 
For apparent $(V-I)$ colours, we first used photometric transformations from \citet{pancino2022} to derive $V$ and $I$ band magnitudes using {\it Gaia} photometric data. These photometric transformations based on homogeneous {\it Gaia} data have been used to provide accurate magnitudes and colours in Johnson-Kron-Cousins photometric systems \citep{trentin2023} and were preferred over the heterogeneous literature compilations. The apparent $V$ and $I$ magnitudes were obtained for Cepheids
which have $(BP-RP)$ colours within the recommended range for photometric transformations \citep{pancino2022}. The colour-excess values in $(V-I)$ were converted using $E(V-I)=1.28E(B-V)$ 
\citep{tammann2003}. Therefore, reddening $E(B-V)$ values were obtained for 76 Cepheids, which vary between 0.092 and  2.186 mag. The uncertainty in the empirical equation used to obtain
intrinsic colours of Cepheids were propagated to the errors in the reddening values.

Recently, \citet{breuval2022} used \texttt{Bayestar19}{\footnote{\url{http://argonaut.skymaps.info/usage}}} three-dimensional reddening maps from \citet{green2019}, and 
period-colour relations from \citet{riess2022} to derive reddening 
values for 222 Cepheids. There is only one star in common with \citet{breuval2022} sample. We also obtained $E(B-V)_\textrm{G19}$ reddening values for 70 Cepheids from \texttt{Bayestar19} maps which 
varied between 0.145 and 4.017 mag. A comparison of the two sets of reddening values for 67 Cepheids suggests a median difference of $E(B-V)_\textrm{G19} - E(B-V) = -0.02$~mag with a scatter of 0.10 mag after 
excluding six stars with the largest differences. If we scale the \texttt{Bayestar19}  reddening values by a scale factor of 0.884 as recommended by \citet{green2019}, the median difference increases to
$-0.11$ mag. Recently, \citet{narloch2023} found that the  \texttt{Bayestar19} reddening maps are inadequate for Cepheids in their sample because the reddening values resulted in 
unexpectedly large scatter in the resulting PL relations. Therefore, we adopted the reddening values based on period-colour relations. 

To correct the magnitudes for extinction, we adopted the \citet{fitzpatrick1999} reddening law assuming an $R_V=3.1$. The total-to-selective absorption ratios were estimated 
using the \texttt{dust\_extinction}{\footnote{\url{https://dust-extinction.readthedocs.io/en/stable/}}} python package. The effective central
wavelength corresponding to each filter was adopted from the Spanish Virtual Observatory filter profile service{\footnote{\url{http://svo2.cab.inta-csic.es/theory/fps/}}}. The absorption
ratios in different filters are provided in Table~\ref{tbl:r_wave}. These absorption ratios were used together with the $E(B-V)$ values to apply extinction corrections to the mean-magnitudes
at all wavelengths. 

\begin{table}
\begin{center}
\caption{Total-to-selective absorption ratios ($R_\lambda$) in different photometric filters for \citet{fitzpatrick1999} reddening law assuming an $R_V=3.1$. \label{tbl:r_wave}}
\small
\begin{tabular}{ccc}
\hline\hline
        {Filter}  & {$\lambda^\textrm{eff}_0$} & $R_\lambda$ \\
\hline

        $G$     &       0.582   &       2.802   \\
        $BP$    &       0.504   &       3.442   \\
        $RP$    &       0.762   &       1.859   \\
        $g$     &       0.467   &       3.804   \\
        $r$     &       0.614   &       2.600   \\
        $i$     &       0.746   &       1.926   \\
        $z$     &       0.892   &       1.423   \\
        $J$     &       1.235   &       0.812   \\
        $H$     &       1.662   &       0.508   \\
        $K_s$   &       2.159   &       0.349   \\
\hline
        \multicolumn{3}{c}{Wesenheit magnitudes}\\
\hline
        $W_{G}$         & \multicolumn{2}{c}{$G-1.900(BP-RP)$} \\       
        $W_{gr}$        & \multicolumn{2}{c}{$r-2.161(g-r)$} \\ 
        $W_{ri}$        & \multicolumn{2}{c}{$i-2.855(r-i)$} \\ 
        $W_{gi}$        & \multicolumn{2}{c}{$i-1.025(g-i)$} \\ 
        $W_{iz}$        & \multicolumn{2}{c}{$z-2.827(i-z)$} \\ 
        $W_{JH}$        & \multicolumn{2}{c}{$H-1.667(J-H)$} \\ 
        $W_{iK_s}$      & \multicolumn{2}{c}{$K_s-0.221(i-K_s)$} \\     
        $W_{JK_s}$      & \multicolumn{2}{c}{$K_s-0.752(J-K_s)$} \\     
\hline
\end{tabular}
\end{center}
\end{table}

\subsection{Period--luminosity relations}
\label{sec:plr_fit}

\begin{table}
\begin{center}
\caption{Multiband period--luminosity relations for MW Cepheids. \label{tbl:plrs}}
\begin{tabular}{ccccc}
\hline\hline
        {Band} & {$\alpha_\lambda$} & {$\beta_\lambda$} & {$\sigma_\textrm{ABL}$}& {$N$}\\
\hline
\multicolumn{5}{c}{All (FU+FO) Cepheids ($\log P_0=1.0$ days)}\\
\hline
              $G$ &$    -4.98\pm0.07    $&$    -3.30\pm0.10    $&     0.04&  71\\
             $BP$ &$    -4.42\pm0.08    $&$    -2.88\pm0.11    $&     0.04&  70\\
             $RP$ &$    -4.95\pm0.05    $&$    -2.84\pm0.08    $&     0.03&  70\\
              $g$ &$    -4.29\pm0.08    $&$    -2.78\pm0.12    $&     0.04&  68\\
              $r$ &$    -4.40\pm0.07    $&$    -2.62\pm0.10    $&     0.04&  70\\
              $i$ &$    -4.60\pm0.06    $&$    -2.73\pm0.09    $&     0.04&  70\\
              $z$ &$    -4.63\pm0.05    $&$    -2.73\pm0.08    $&     0.04&  73\\
              $J$ &$    -5.65\pm0.04    $&$    -2.99\pm0.07    $&     0.02&  68\\
              $H$ &$    -6.02\pm0.04    $&$    -3.23\pm0.07    $&     0.02&  70\\
            $K_s$ &$    -6.06\pm0.04    $&$    -3.26\pm0.07    $&     0.02&  69\\
\hline
\multicolumn{5}{c}{Only FU Cepheids ($\log P_0=1.0$ days)}\\
\hline
              $G$ &$    -5.06\pm0.06    $&$    -3.12\pm0.14    $&     0.04&  43\\
             $BP$ &$    -4.47\pm0.07    $&$    -2.70\pm0.16    $&     0.04&  43\\
             $RP$ &$    -4.98\pm0.05    $&$    -2.74\pm0.11    $&     0.04&  45\\
              $g$ &$    -4.30\pm0.09    $&$    -2.81\pm0.18    $&     0.04&  41\\
              $r$ &$    -4.44\pm0.07    $&$    -2.50\pm0.14    $&     0.04&  43\\
              $i$ &$    -4.62\pm0.05    $&$    -2.61\pm0.12    $&     0.04&  45\\
              $z$ &$    -4.65\pm0.05    $&$    -2.55\pm0.11    $&     0.05&  45\\
              $J$ &$    -5.67\pm0.04    $&$    -2.85\pm0.10    $&     0.02&  42\\
              $H$ &$    -6.03\pm0.04    $&$    -3.03\pm0.10    $&     0.02&  44\\
            $K_s$ &$    -6.05\pm0.04    $&$    -3.10\pm0.10    $&     0.02&  44\\
\hline
\multicolumn{5}{c}{Only FO Cepheids ($\log P_0=0.4$ days)}\\
\hline
              $G$ &$    -3.42\pm0.07    $&$    -3.01\pm0.28    $&     0.04&  27\\
             $BP$ &$    -3.06\pm0.08    $&$    -2.66\pm0.32    $&     0.04&  27\\
             $RP$ &$    -3.64\pm0.05    $&$    -2.80\pm0.22    $&     0.03&  27\\
              $g$ &$    -3.02\pm0.09    $&$    -2.52\pm0.35    $&     0.04&  27\\
              $r$ &$    -3.18\pm0.07    $&$    -2.58\pm0.27    $&     0.04&  27\\
              $i$ &$    -3.33\pm0.06    $&$    -2.80\pm0.22    $&     0.04&  27\\
              $z$ &$    -3.38\pm0.05    $&$    -3.05\pm0.19    $&     0.04&  28\\
              $J$ &$    -4.27\pm0.05    $&$    -3.21\pm0.17    $&     0.03&  27\\
              $H$ &$    -4.54\pm0.04    $&$    -3.47\pm0.15    $&     0.03&  28\\
            $K_s$ &$    -4.57\pm0.04    $&$    -3.44\pm0.15    $&     0.02&  26\\
\hline
\end{tabular}
\end{center}
        \footnotesize{{\bf Notes:} The zero-point ($\alpha$), slope ($\beta$), dispersion ($\sigma_\textrm{ABL}$) of ABL fits, and 
the number of stars ($N$) in the final PL relations are listed.}
\end{table}

\begin{table}
\begin{center}
\caption{Period-Wesenheit relations for MW Cepheids. \label{tbl:pwrs}}
\begin{tabular}{ccccc}
\hline\hline
        {Band} & {$\alpha_\lambda$} & {$\beta_\lambda$} & {$\sigma_\textrm{ABL}$}& {$N$}\\
\hline
\multicolumn{5}{c}{All (FU+FO) Cepheids ($\log P_0=1.0$ days)}\\
\hline
            $W_G$ &$    -6.27\pm0.04    $&$    -3.54\pm0.06    $&     0.02&  74\\
         $W_{gr}$ &$    -5.08\pm0.06    $&$    -3.10\pm0.10    $&     0.04&  68\\
         $W_{ri}$ &$    -5.24\pm0.06    $&$    -3.36\pm0.09    $&     0.03&  75\\
         $W_{gi}$ &$    -4.99\pm0.05    $&$    -2.93\pm0.07    $&     0.04&  73\\
         $W_{iz}$ &$    -4.76\pm0.07    $&$    -2.87\pm0.10    $&     0.04&  77\\
         $W_{JH}$ &$    -6.61\pm0.06    $&$    -3.60\pm0.09    $&     0.01&  74\\
         $W_{iK}$ &$    -6.38\pm0.04    $&$    -3.35\pm0.06    $&     0.01&  71\\
         $W_{JK}$ &$    -6.35\pm0.04    $&$    -3.42\pm0.07    $&     0.01&  73\\
\hline
\multicolumn{5}{c}{Only FU Cepheids ($\log P_0=1.0$ days)}\\
\hline
            $W_G$ &$    -6.31\pm0.04    $&$    -3.44\pm0.09    $&     0.01&  45\\
         $W_{gr}$ &$    -5.14\pm0.06    $&$    -2.35\pm0.13    $&     0.05&  44\\
         $W_{ri}$ &$    -5.28\pm0.07    $&$    -3.18\pm0.14    $&     0.02&  44\\
         $W_{gi}$ &$    -5.04\pm0.04    $&$    -2.63\pm0.10    $&     0.04&  45\\
         $W_{iz}$ &$    -4.77\pm0.07    $&$    -2.61\pm0.13    $&     0.05&  49\\
         $W_{JH}$ &$    -6.64\pm0.06    $&$    -3.30\pm0.13    $&     0.01&  45\\
         $W_{iK}$ &$    -6.38\pm0.04    $&$    -3.17\pm0.09    $&     0.02&  48\\
         $W_{JK}$ &$    -6.36\pm0.05    $&$    -3.22\pm0.10    $&     0.02&  48\\
\hline
\multicolumn{5}{c}{Only FO Cepheids ($\log P_0=0.4$ days)}\\
\hline
            $W_G$ &$    -4.65\pm0.03    $&$    -3.45\pm0.15    $&     0.02&  27\\
         $W_{gr}$ &$    -3.48\pm0.06    $&$    -2.99\pm0.19    $&     0.04&  29\\
         $W_{ri}$ &$    -3.70\pm0.06    $&$    -3.53\pm0.21    $&     0.04&  28\\
         $W_{gi}$ &$    -3.62\pm0.04    $&$    -3.19\pm0.16    $&     0.04&  28\\
         $W_{iz}$ &$    -3.48\pm0.06    $&$    -3.50\pm0.22    $&     0.03&  26\\
         $W_{JH}$ &$    -4.95\pm0.05    $&$    -3.85\pm0.20    $&     0.02&  27\\
         $W_{iK}$ &$    -4.85\pm0.03    $&$    -3.59\pm0.14    $&     0.01&  26\\
         $W_{JK}$ &$    -4.79\pm0.04    $&$    -3.65\pm0.15    $&     0.02&  27\\
\hline
\end{tabular}
\end{center}
        \footnotesize{{\bf Notes:} The zero-point ($\alpha$), slope ($\beta$), dispersion ($\sigma_\textrm{ABL}$) of ABL fits, and 
the number of stars ($N$) in the final PW relations are listed.}
\end{table}

The parallax uncertainties for Cepheids in our sample are large ($>25\%$) for 17 stars and two stars have negative parallaxes. Moreover, 12 Cepheids have distances larger than 10 kpc. We did not
adopt distances from \citet{bailer2021} for which priors based on a three-dimensional model of the Galaxy become dominant for these distant targets. Therefore, instead of converting apparent magnitudes
to absolute magnitudes using geometric distances, we decided to work in the parallax space to derive PL relations. We derived astrometry-based luminosity \citep[ABL,][]{feast1997,arenou1999} defined as:
\begin{eqnarray}
 {\textrm{ABL}} ~&=&~ \overline{\omega}_\textrm{(mas)}10^{0.2m_\lambda-2} \nonumber \\
        ~&=&~ 10^{0.2(\alpha_{\lambda} + \beta_{\lambda} (\log P_{i} - \log P_0))},
    \label{eq:plr_abl}
\end{eqnarray}

\noindent where $m_\lambda$ is the extinction corrected magnitude at a given wavelength for a Cepheid with period ($P_i$). The period ($P_0$) at which the zero-point is determined, is adopted at 
$\log P_0=1.0$ days for FU Cepheids and $\log P_0=0.4$ days for FO Cepheids. Since our sample size of each subtype is small, we also considered a combined sample of Cepheids by fundamentalizing the 
periods of FO Cepheids using the equation: $P_{FU} = P_{FO}/(0.716-0.027 \log P_{FO})$ from \citet{feast1997}. The absolute zero-point was also obtained at $\log P_0=1.0$ days for this combined sample. 
While fitting these relations in the form of equation~(\ref{eq:plr_abl}),
we iteratively removed the single largest outlier in each iteration until all residuals are within $\pm 3\sigma$, where $\sigma$ represents the root-mean square (rms) error. We created $10^4$ random realizations of ABL fits to derive coefficients and their associated uncertainties. In this procedure, we also included the parallax zero-point offset error of $6~\mu$as \citep{riess2021}.

\begin{figure*}
  \centering
  \begin{tabular}{ccc}
    \includegraphics[width=0.315\textwidth]{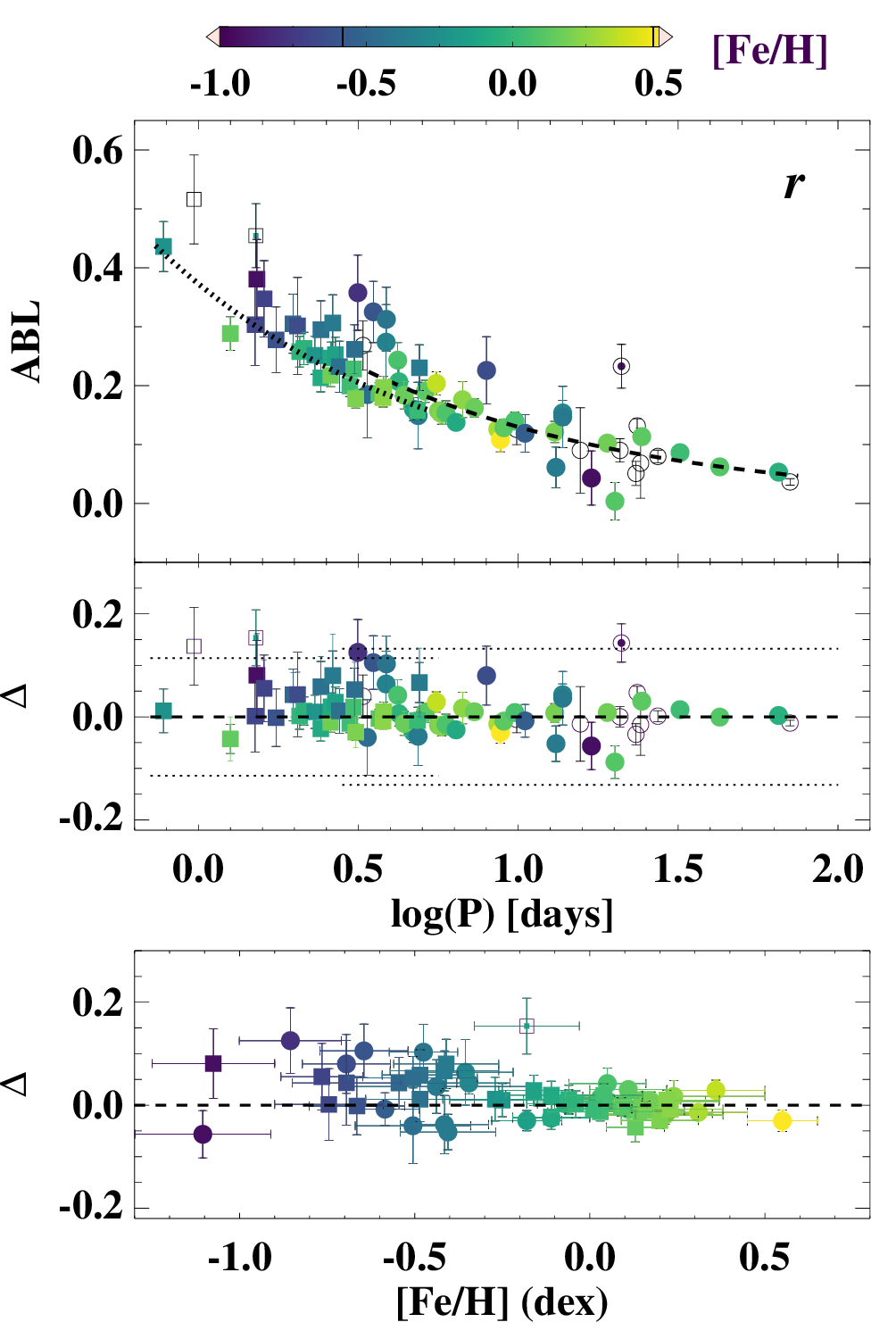} & 
      \includegraphics[width=0.315\textwidth]{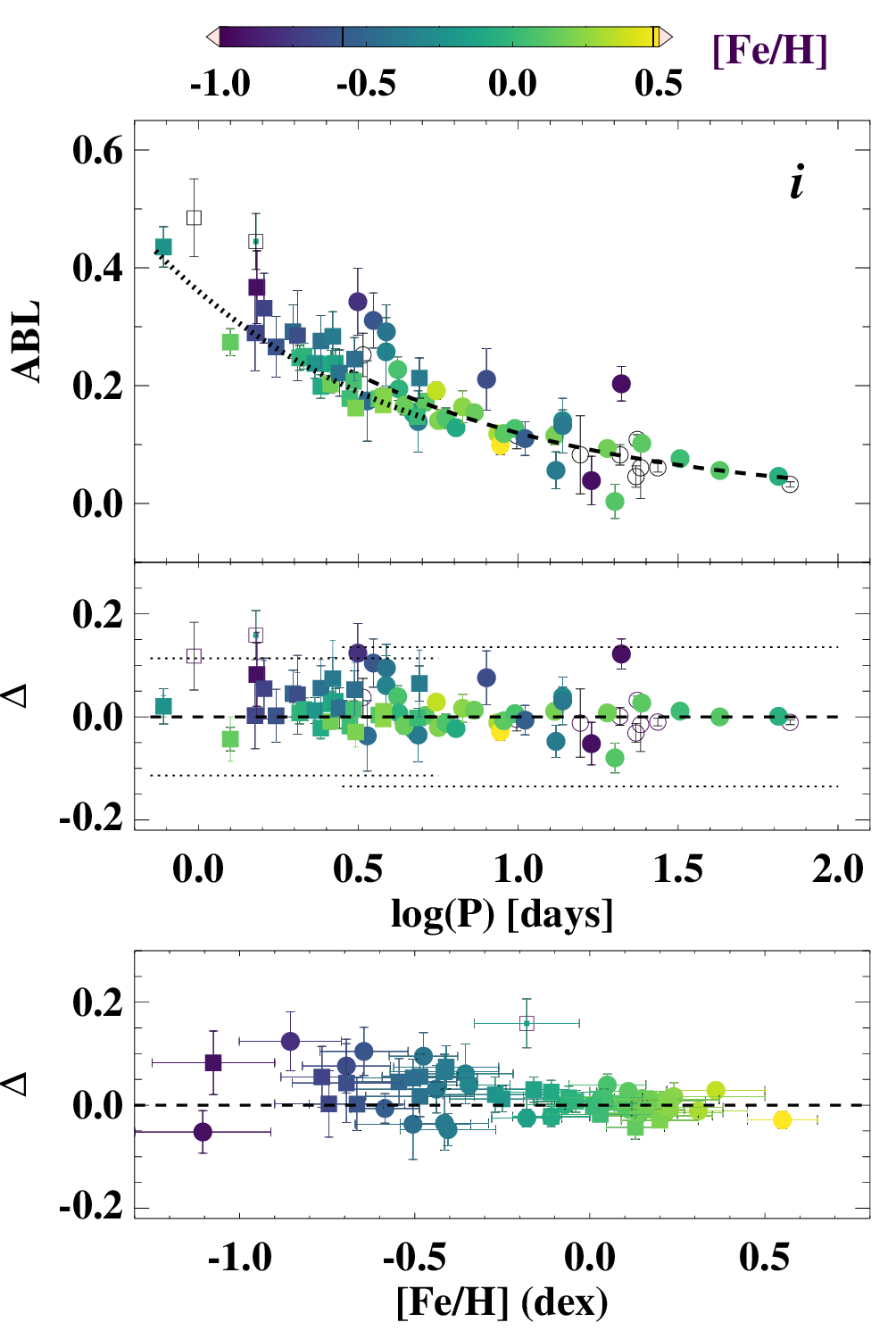}&
  \includegraphics[width=0.315\textwidth]{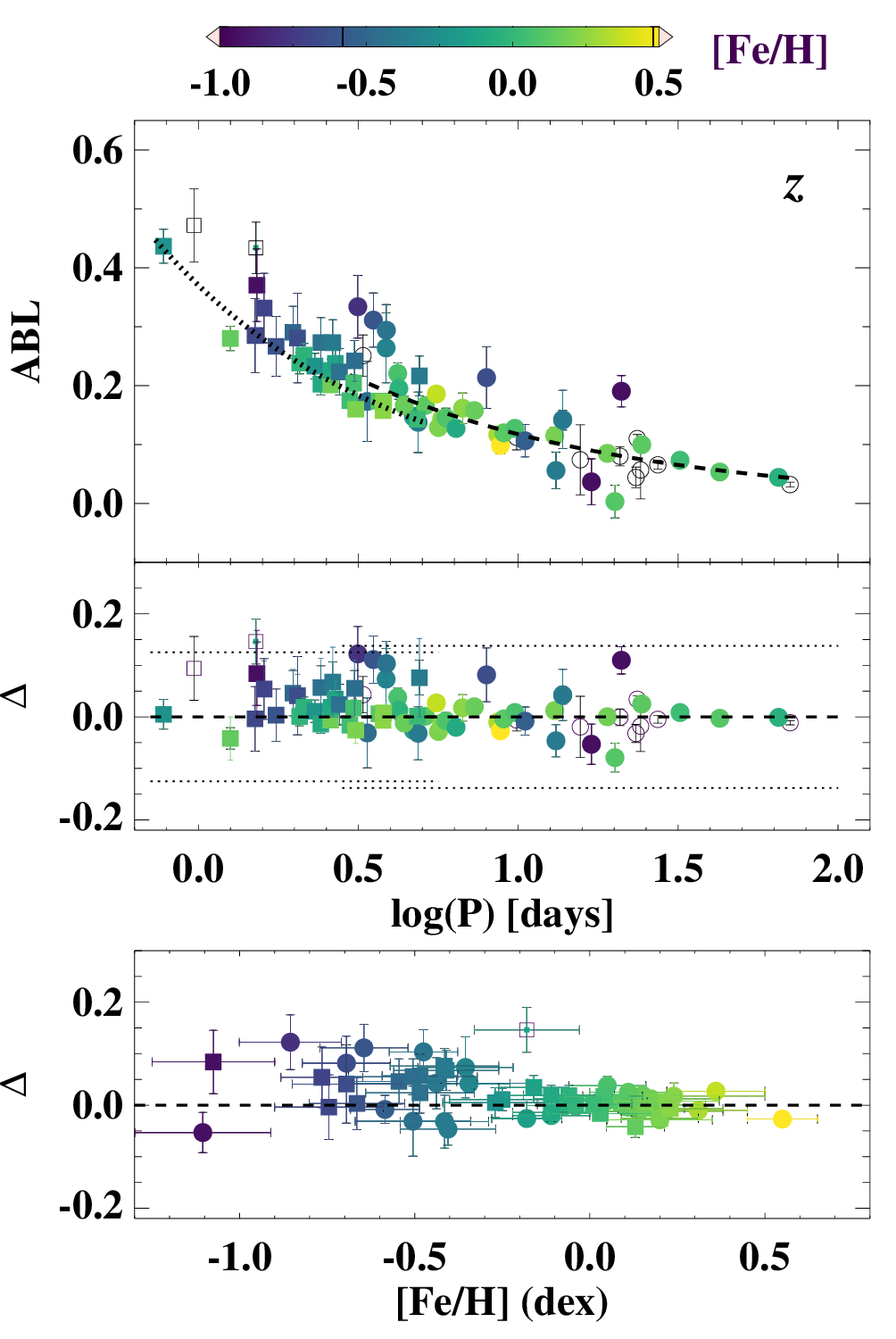} \\
    \includegraphics[width=0.315\textwidth]{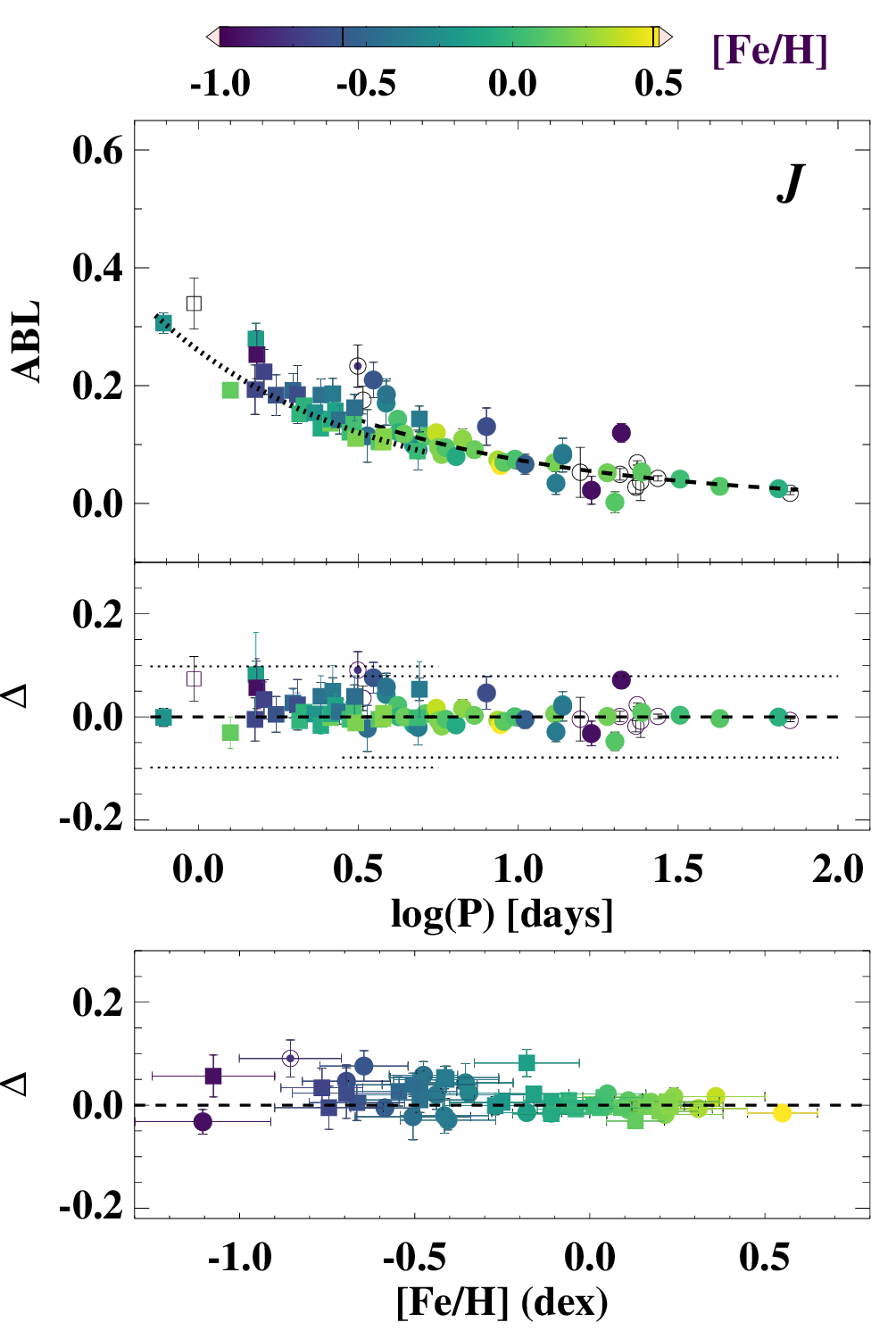} & 
      \includegraphics[width=0.315\textwidth]{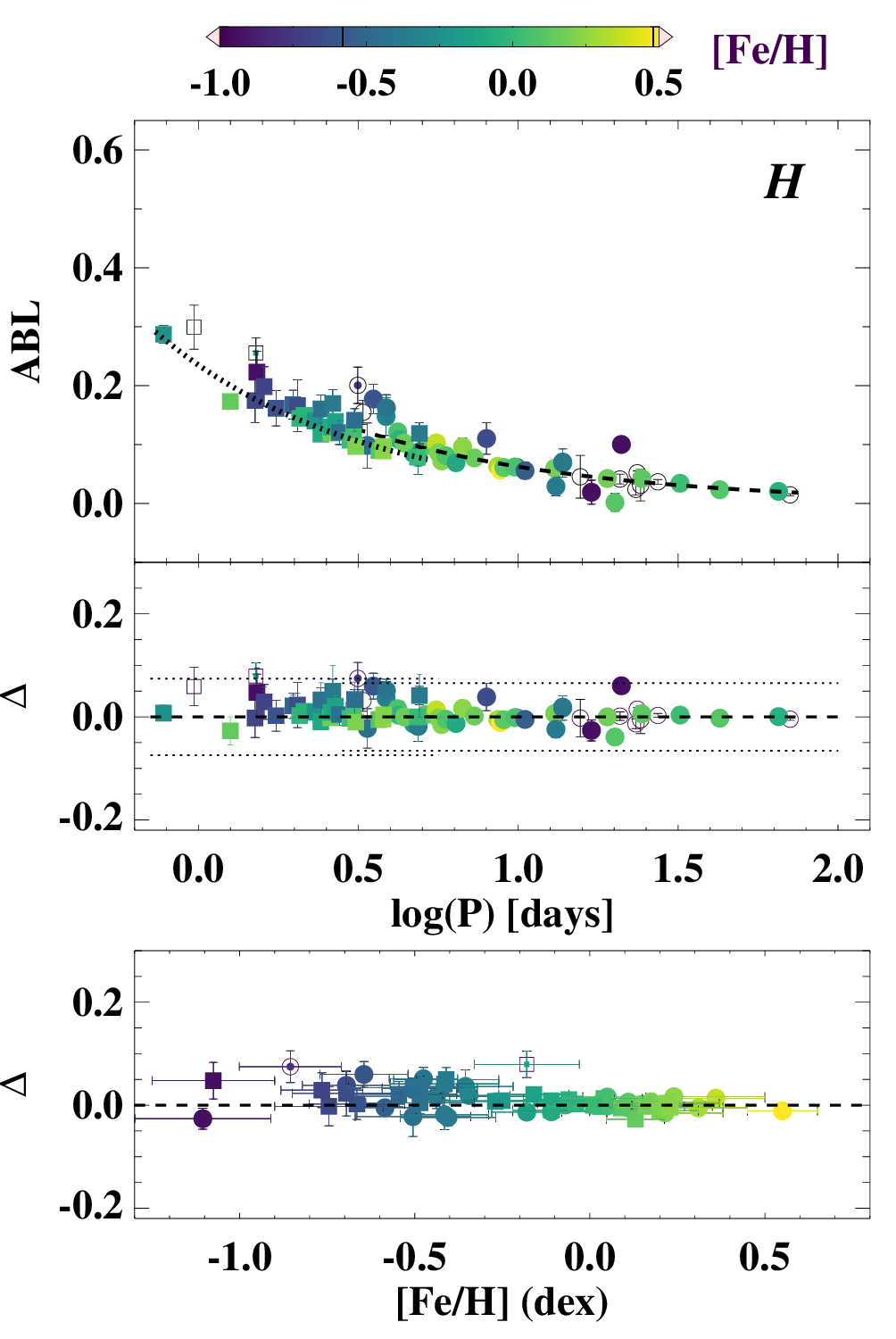}&
  \includegraphics[width=0.315\textwidth]{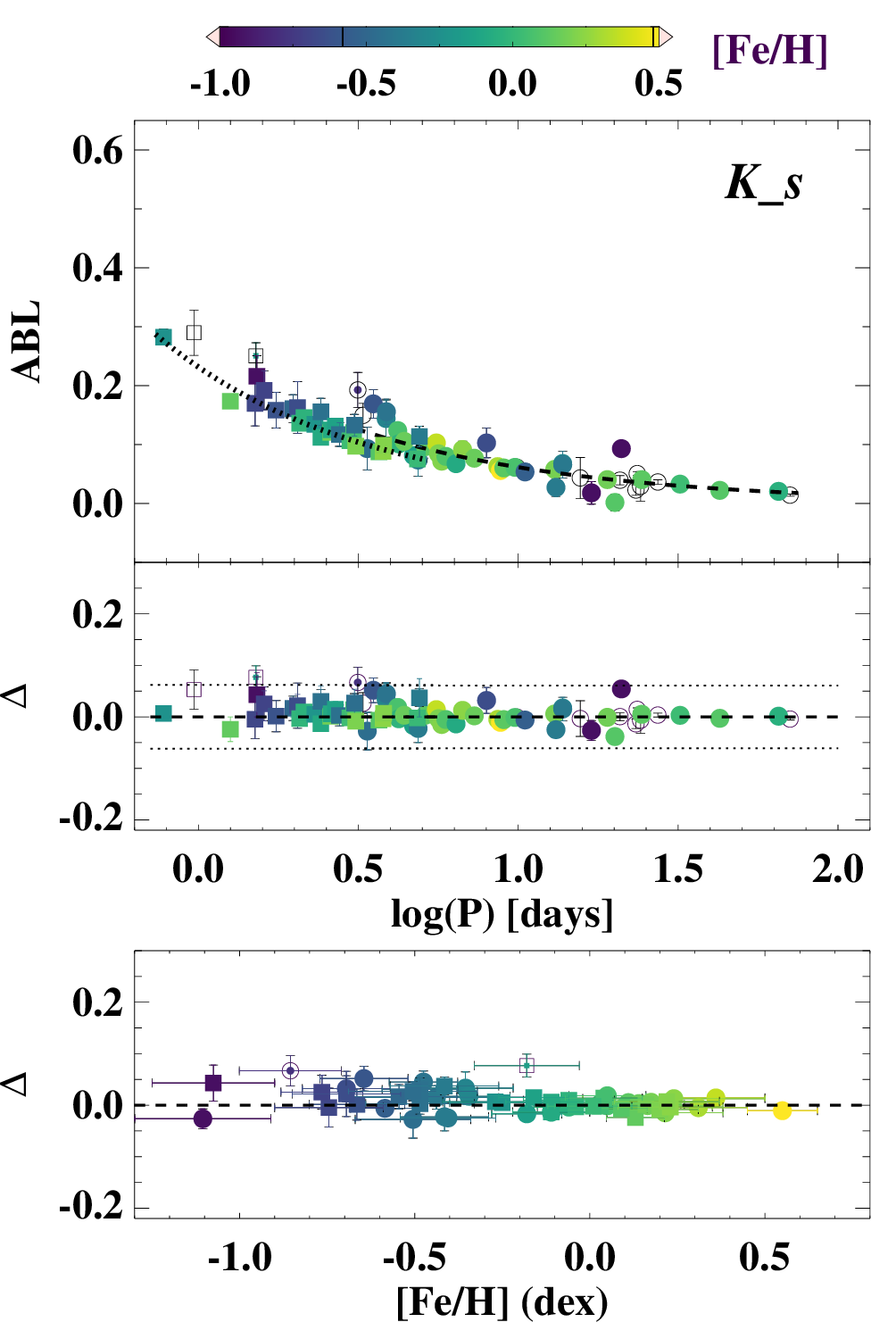} \\
\end{tabular}
        \caption{Astrometry-based luminosity of FU (circles) and FO (squares) Cepheids as a function of pulsation period with magnitudes in $rizJHK_s$ bands is shown in the   top part of each subpanel. The dashed and dotted lines represent the best-fitting relation for FU and FO mode Cepheids, respectively. The middle plot in each subpanel displays the residuals of the best fit as a function of period. The dashed lines show $\pm3\sigma$ scatter for FU and FO modes Cepheids. The bottom plot in each subpanel shows the residuals as a function of metallicity, if available. The dashed lines in the bottom two subpanels represent zero-residual variation. The partially filled symbols represent outliers, while empty symbols represent no metallicity measurement. The colour bar represents metallicity. }
             \label{fig:abl_plr}
\end{figure*}

\begin{figure*}
  \centering
  \begin{tabular}{ccc}
      \includegraphics[width=0.3\textwidth]{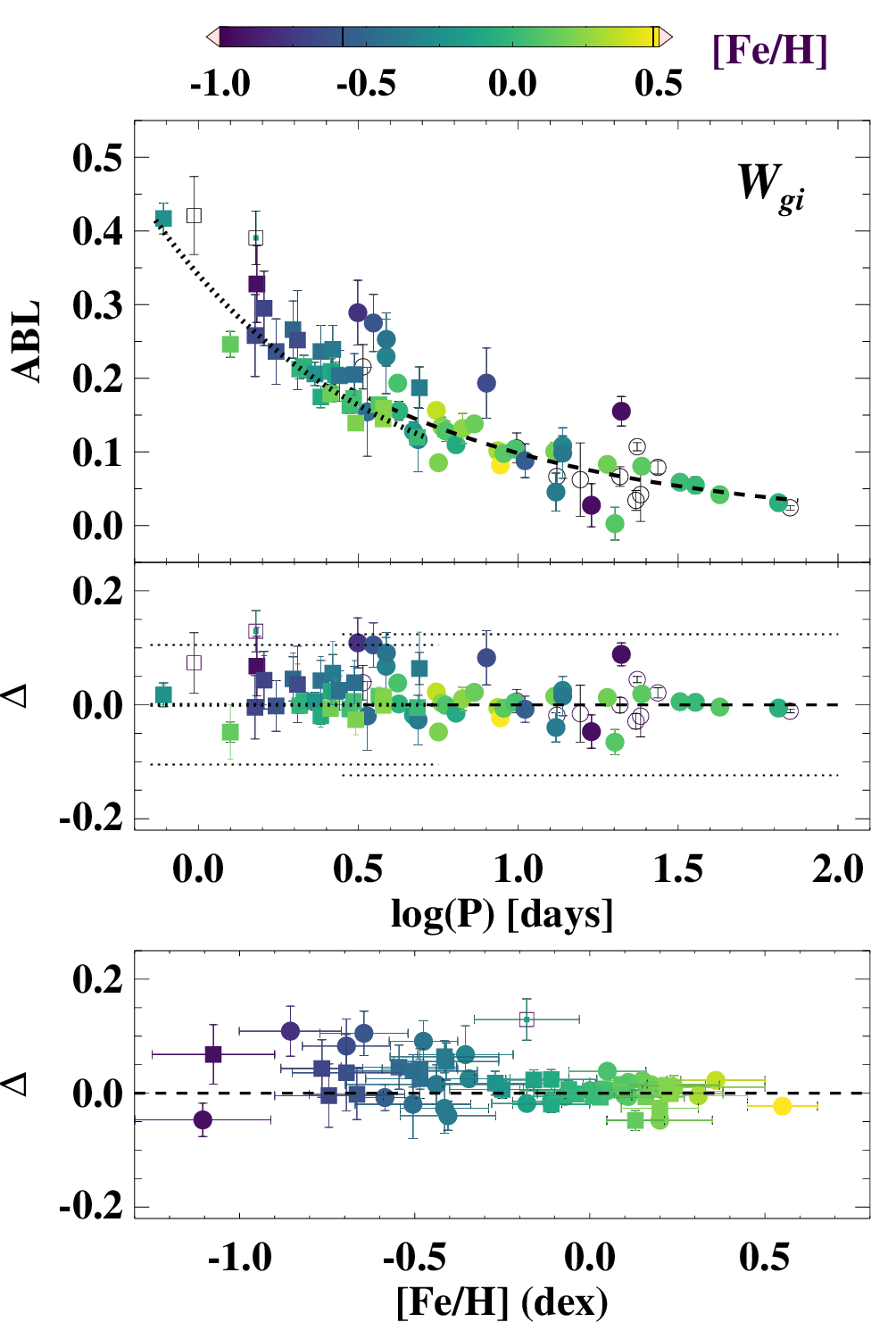} &  
      \includegraphics[width=0.3\textwidth]{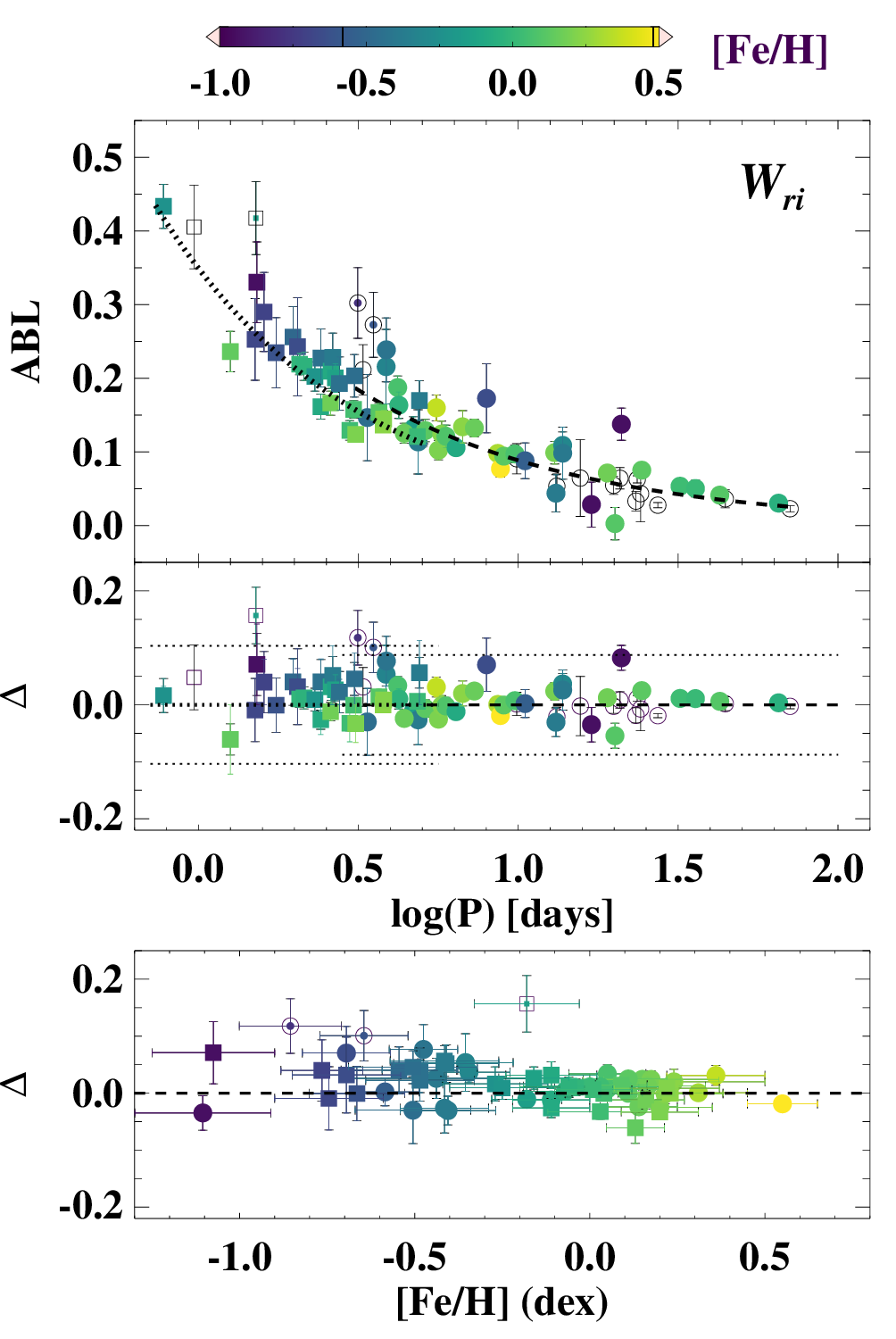}&
  \includegraphics[width=0.3\textwidth]{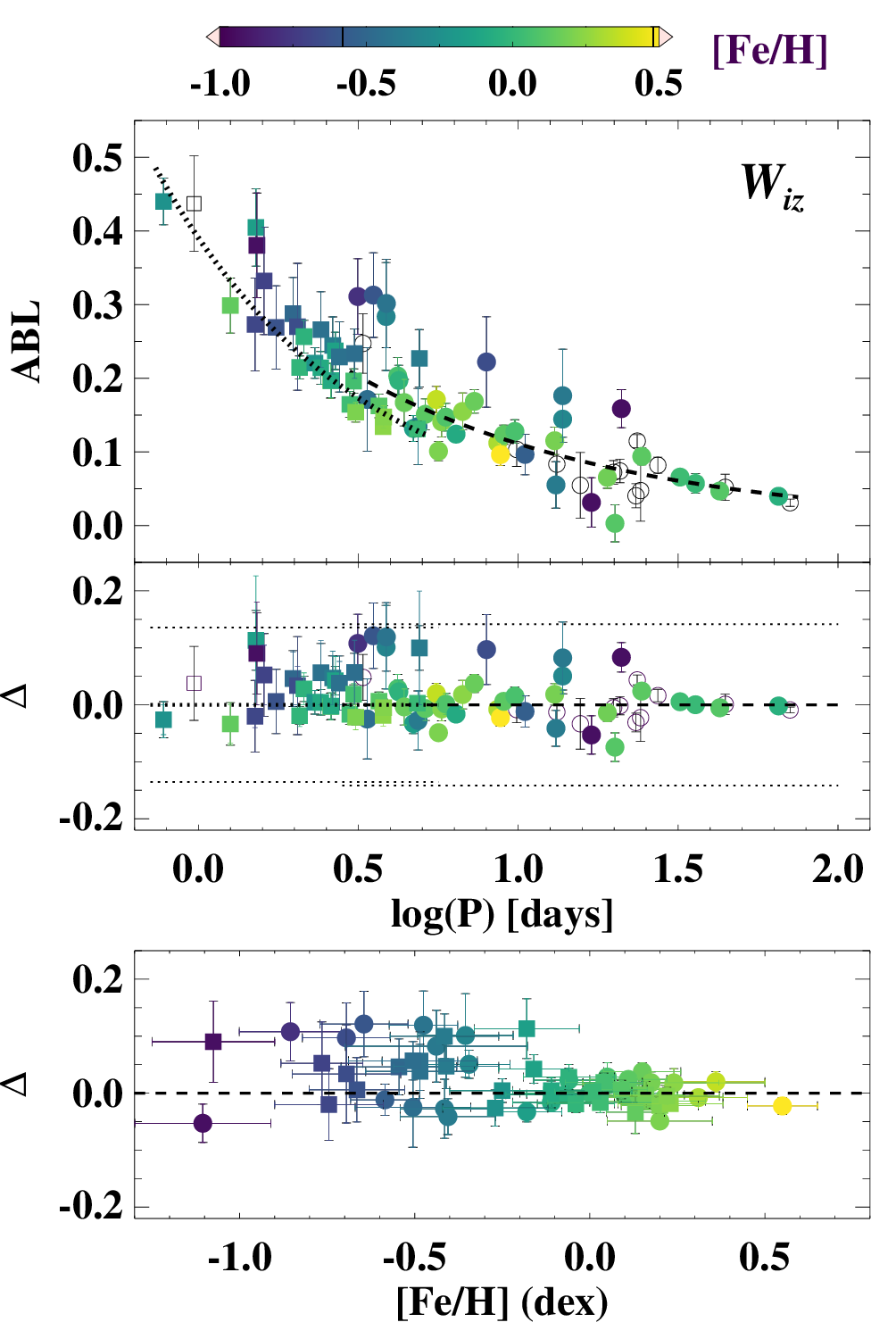} \\
      \includegraphics[width=0.3\textwidth]{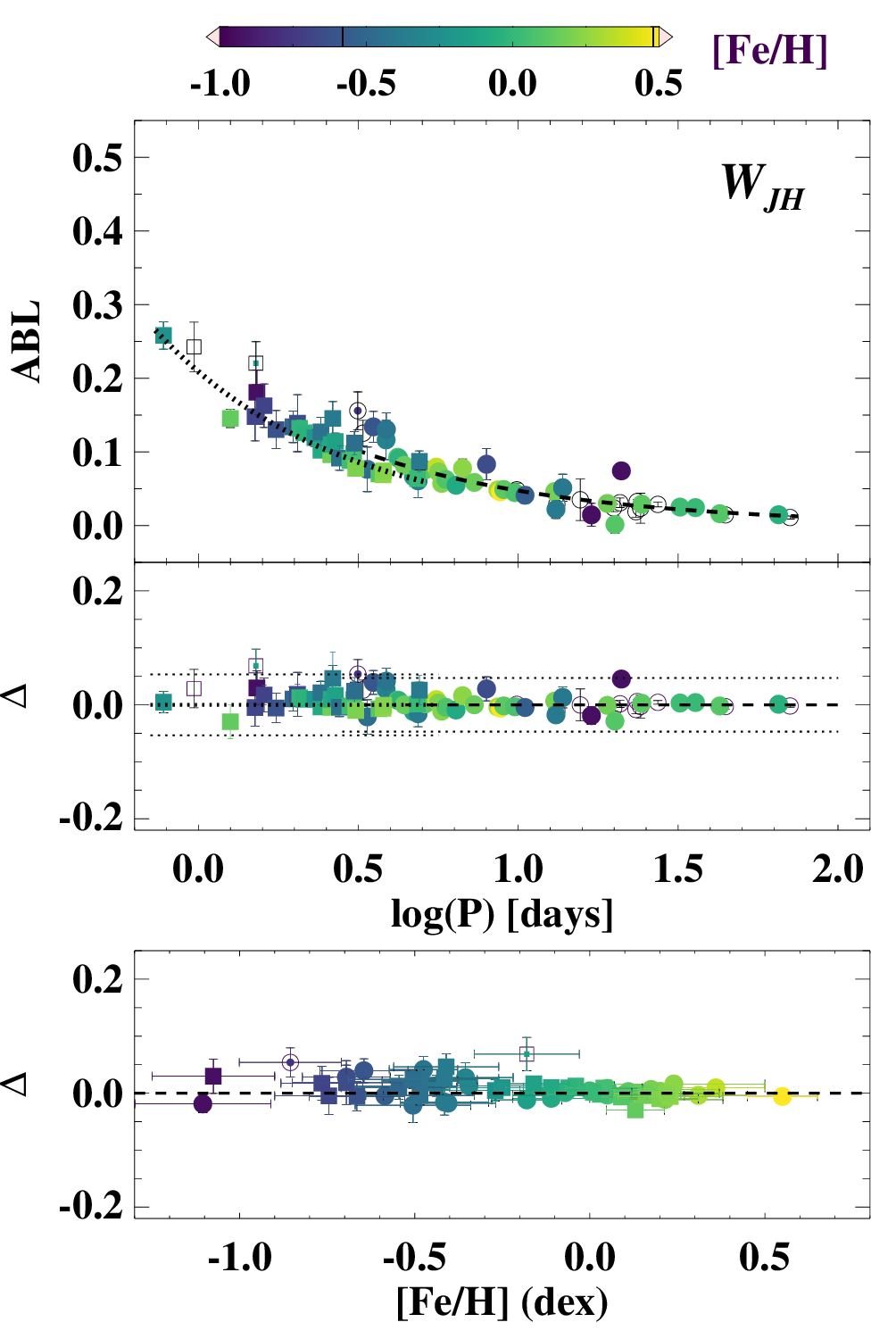} &  
      \includegraphics[width=0.3\textwidth]{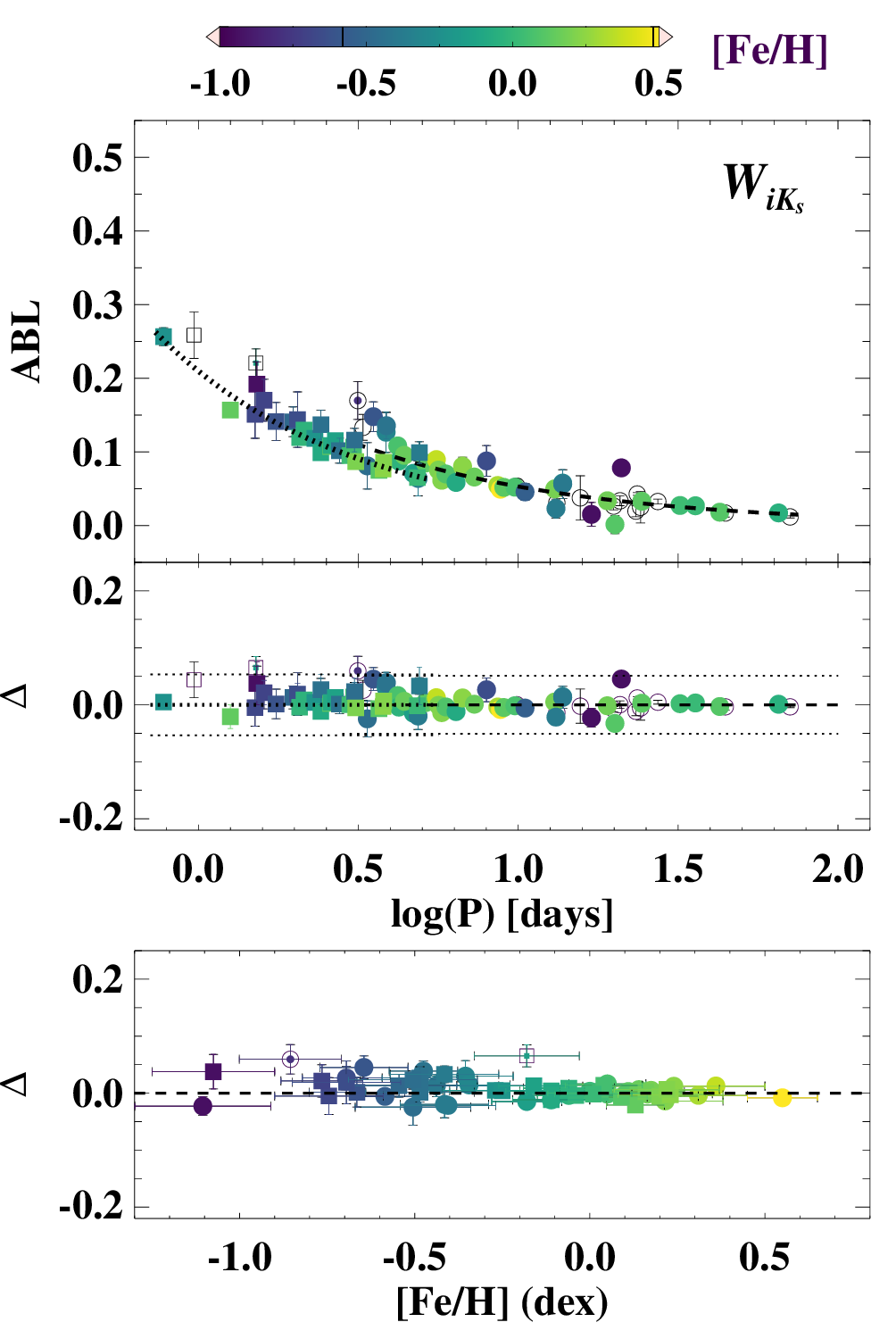}&
  \includegraphics[width=0.3\textwidth]{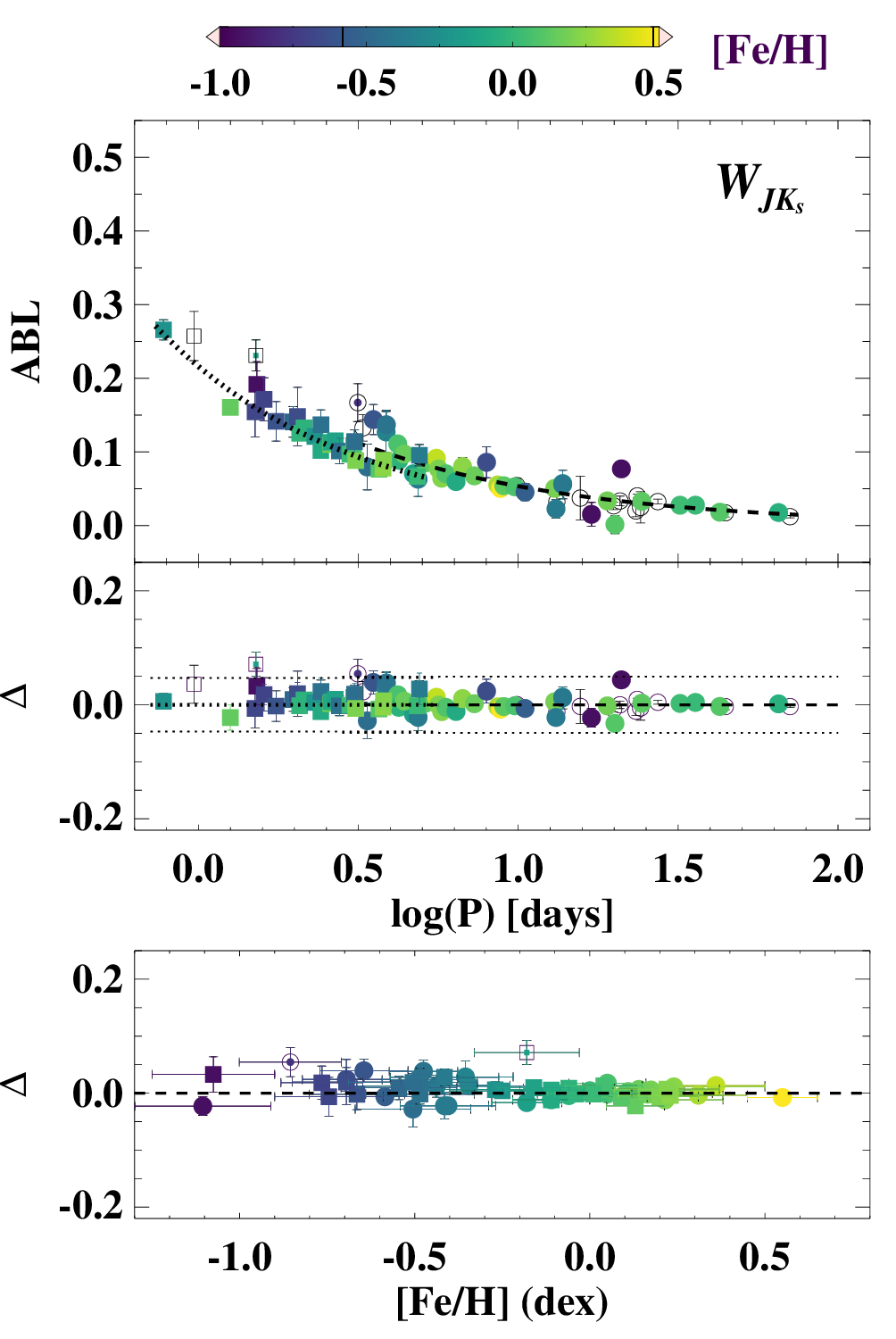} \\
\end{tabular}
        \caption{Same as Figure~\ref{fig:abl_plr},  but for Wesenheit magnitudes.}
             \label{fig:abl_pwr}
\end{figure*}

Figure~\ref{fig:abl_plr} displays the ABL of Cepheids in multiple bands and the results of the best-fitting relation in the form of equation~(\ref{eq:plr_abl}) are
listed in Table~\ref{tbl:plrs}. A clear decrease in the scatter and PL residuals is seen moving from optical to NIR wavelengths. Among the Cepheids that have metallicity measurements, the residuals
are larger for metal-poor stars ($\textrm{[Fe/H]}<-0.3$ dex). This suggests that the metal-poor stars have higher ABL and fainter absolute magnitudes for a given period. This is consistent with recent
results of a negative metallicity coefficient of PL relations in the literature \citep[e.g.][]{breuval2022, molinaro2023, bhardwaj2023a}. However, the metal-poor stars, most of which are more distant, also 
have larger parallax uncertainties. We note that the PL relation in the $g$ band is not shown in Figure~\ref{fig:abl_plr}, but exhibits trends similar to other optical filters.
The results given in Table~\ref{tbl:plrs} show that the slopes of PL relations are steeper at NIR wavelengths than optical bands, a trend that is typical
for both Cepheid and RR Lyrae stars \citep{bhardwaj2020}. 

\subsection{Period-Wesenheit relations}

The lack of accurate and independent reddening values of Cepheids in our sample can impact their PL relations. The Wesenheit magnitudes, which include a colour-term, are constructed to be reddening
independent \citep{madore1982}. Given a reddening law, the coefficient of the colour-term of a two-band Wesenheit magnitude is derived from the total-to-selective absorption parameters at 
those wavelengths. We adopted the \citet{fitzpatrick1999} reddening law and derived PW relations for different combination of bandpasses as listed in Table~\ref{tbl:r_wave}. 
Figure~\ref{fig:abl_pwr} displays ABL for the PW relations for Cepheids in our sample. These relations are apparently tighter than the PL relations suggesting that the reddening values based on 
period-colour relations are not very accurate. The trend of larger residuals for metal-poor stars can also be seen in the bottom panels similar to PL relations. Table~\ref{tbl:pwrs} lists
the coefficients of PW relations.  

We tested the PL and PW relations listed in Tables~\ref{tbl:plrs} and \ref{tbl:pwrs} for different samples after excluding light curves with poor quality flags, Cepheids with $\textrm{RUWE}>1.4$ and
$\textrm{GOF}>12.5$, and applying different outlier rejection thresholds. No statistically significant variations are seen in the coefficients of the PL/PW relations. If we restrict the sample to
parallax uncertainties smaller than $15\%$ and/or exclude stars beyond 10~kpc, the coefficients of PL relations vary within $1\sigma$ of their quoted uncertainties, which increase with lower statistics. 
While using reddening values from the maps of \citet{green2019}, the PL relations exhibit significantly larger scatter, as noted by \citet{narloch2023}. The dominant source of scatter in our PL and PW relations is due to
large parallax uncertainties, which are expected to improve in the future {\it Gaia} data releases. 

\subsection{Comparison with literature}

\begin{figure*}
\centering
\includegraphics[width=0.85\textwidth]{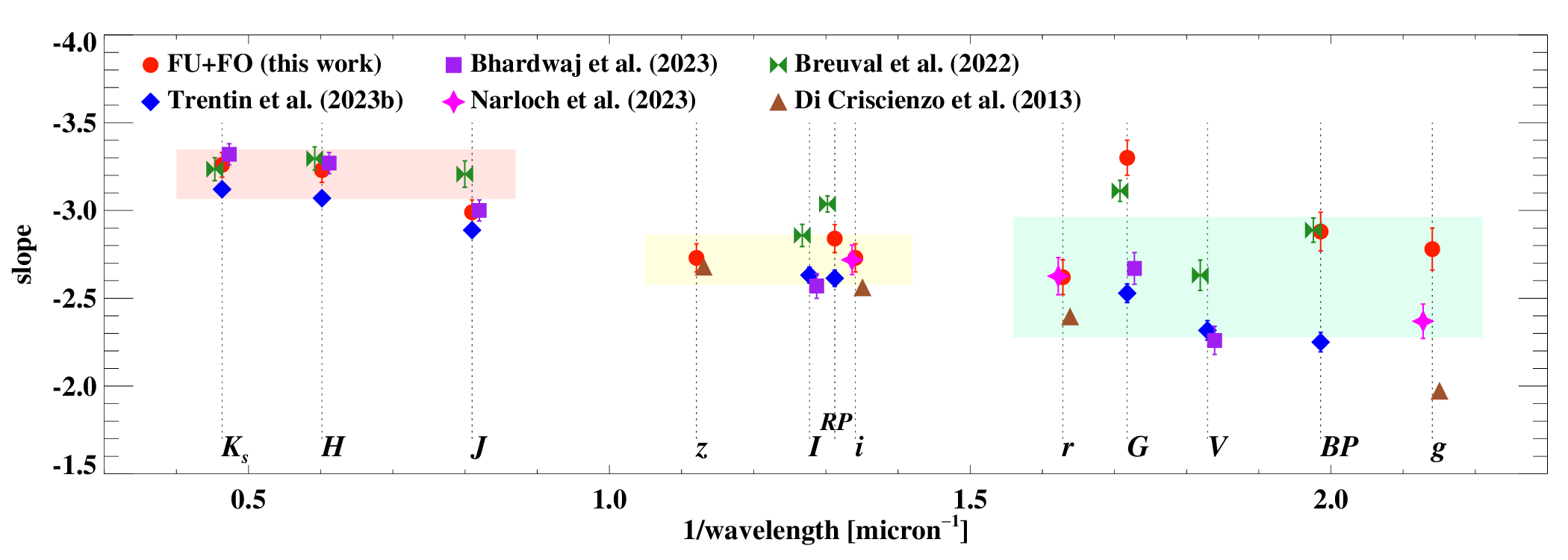}  
        \caption{Comparison of slopes of PL relations at different wavelengths. The slopes of the Sloan band PL relations are compared with empirical ($gri$) and theoretical ($griz$) PL relations from \citet{narloch2023} and \citet{dicriscienzo2013}, respectively. The slopes of the $G, BP, RP, V, I, J, H,$ and $K_s$ band PL relations are from \citet{breuval2022}, \citet{trentin2023a}, and \citet{bhardwaj2023a}. The shaded regions display the  $\pm1 \sigma$ standard deviation around the mean value of the slopes for the wavelengths under consideration.}
        \label{fig:slp_comp}
\end{figure*}

Figure~\ref{fig:slp_comp} displays a comparison of slopes of PL relations in optical and NIR filters with previously reported values in the recent literature. The typical trend of steeper slopes 
at NIR wavelengths is seen. However, we did not find a distinct linear relation between the slopes and the wavelength across all filters, as noted by \citet{trentin2023a}. The slopes of $ri$ band PL relations for the
combined sample of Cepheids are in excellent agreement with the slopes of empirical PL relations for FU Cepheids derived by \citet{narloch2023}. The slope of $g$ band PL relation in this work is steeper than that derived by \citet{narloch2023}, but 
the difference is still within $2\sigma$ of their quoted uncertainties. \citet{dicriscienzo2013} provided theoretical PL relations in Sloan filters based on non-linear pulsation models representative of Cepheids in our Galaxy. The authors found their predicted slopes to be mildly steeper than the empirical slopes of PL relations derived using photometric transformation 
from traditional Johnson-Cousins to Sloan bands. The theoretical $griz$ band slopes in \citet{dicriscienzo2013} are also steeper for shorter period variables ($\log P < 1.0$~days) than those for longer period stars. The slopes of $riz$ band PL relations in this work are in good agreement with the theoretically predicted slopes for the entire period range models, but are significantly steeper in the $g$-band. The slopes of our NIR PL relations for Cepheids are in good agreement with the results of \citet{breuval2022}, \citet{trentin2023a}, and \citet{bhardwaj2023a}. The slopes of $G$, $BP$ and $RP$ band PL relations are in agreement with those from \citet{breuval2022}. In the case of optical PW relations, the adopted filter combinations and/or the color-coefficients are different in this work as compared to \citet{narloch2023}. Nevertheless, the slopes of $W_{gr}$ PW relations are in good agreement when we consider the combined sample of Cepheids. The slope of $W_G$ P-W relation is similar to the one derived by \citet{breuval2022}, while it is consistent within $2\sigma$ for the $W_{JK_s}$ PW relation.

\section{Metallicity dependence of the Leavitt law}
\label{sec:plz}

\begin{figure*}
  \centering
  \begin{tabular}{ccc}
      \includegraphics[width=0.3\textwidth]{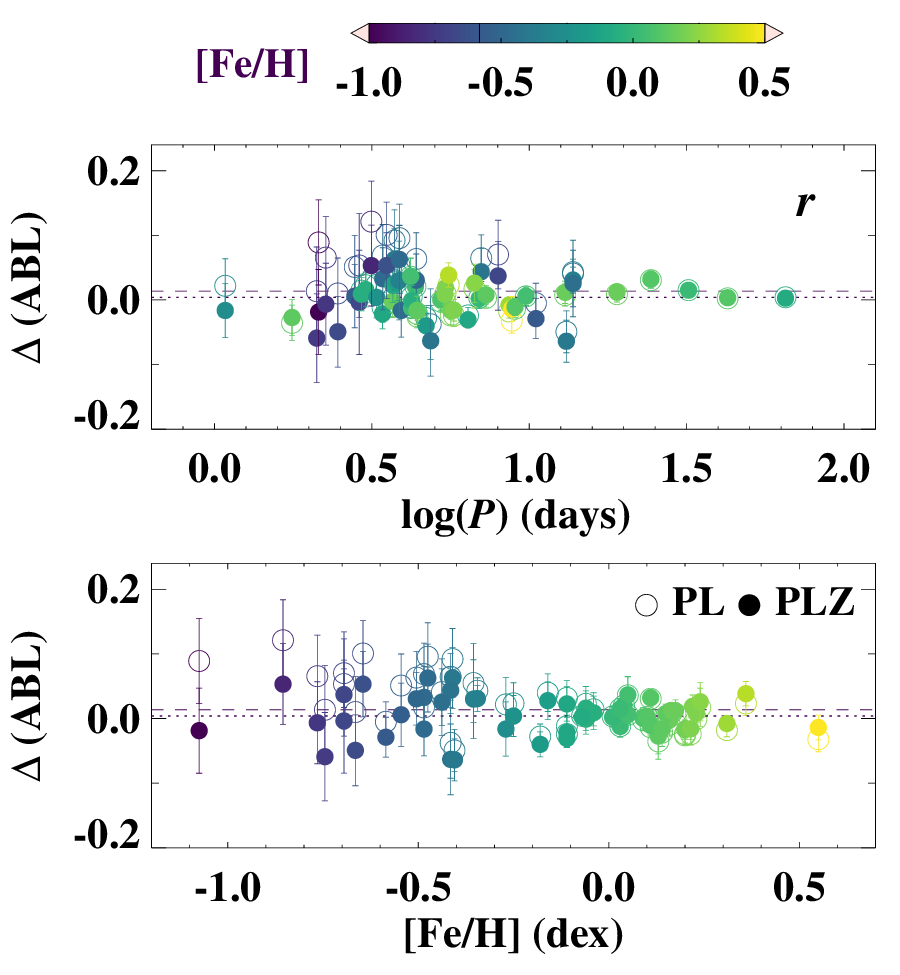} &  
      \includegraphics[width=0.3\textwidth]{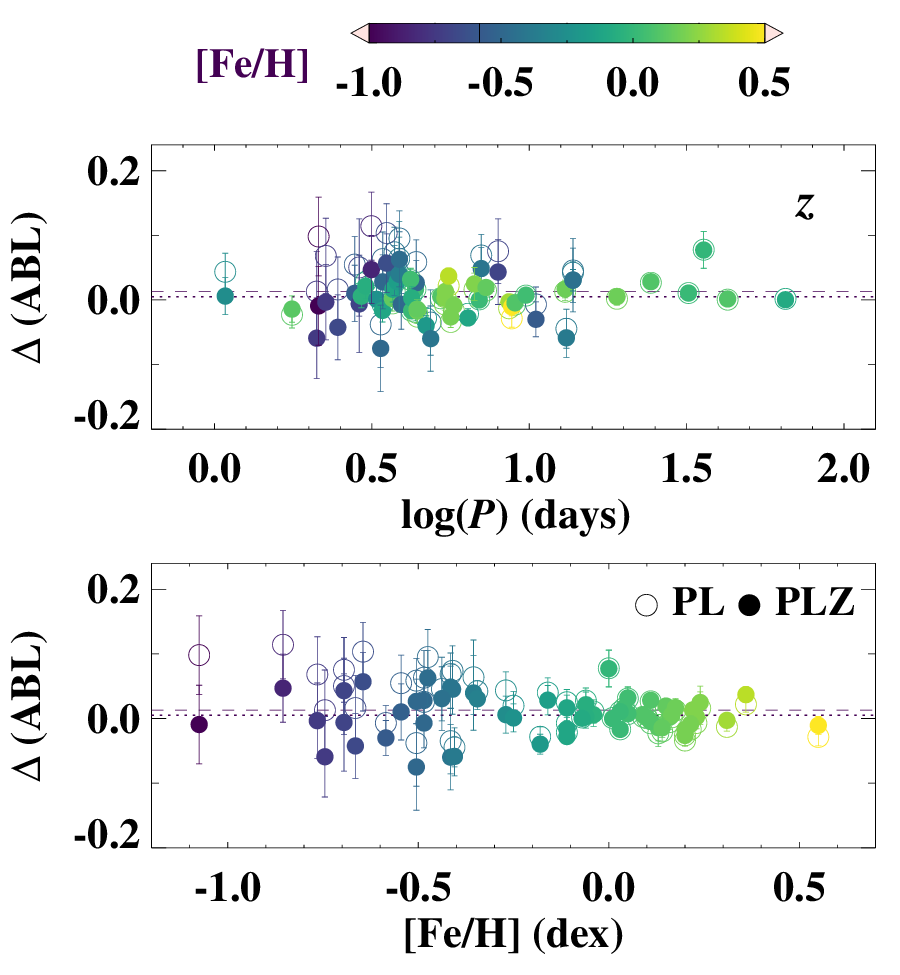} &  
          \includegraphics[width=0.3\textwidth]{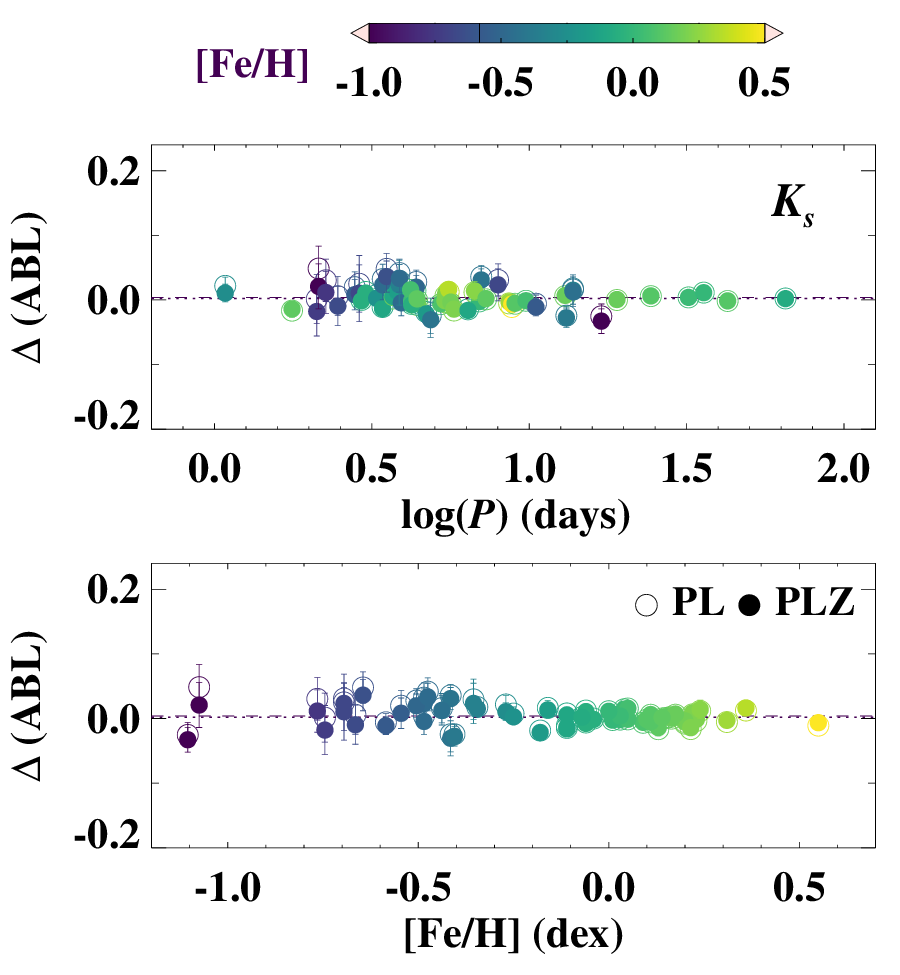} \\  
      \includegraphics[width=0.3\textwidth]{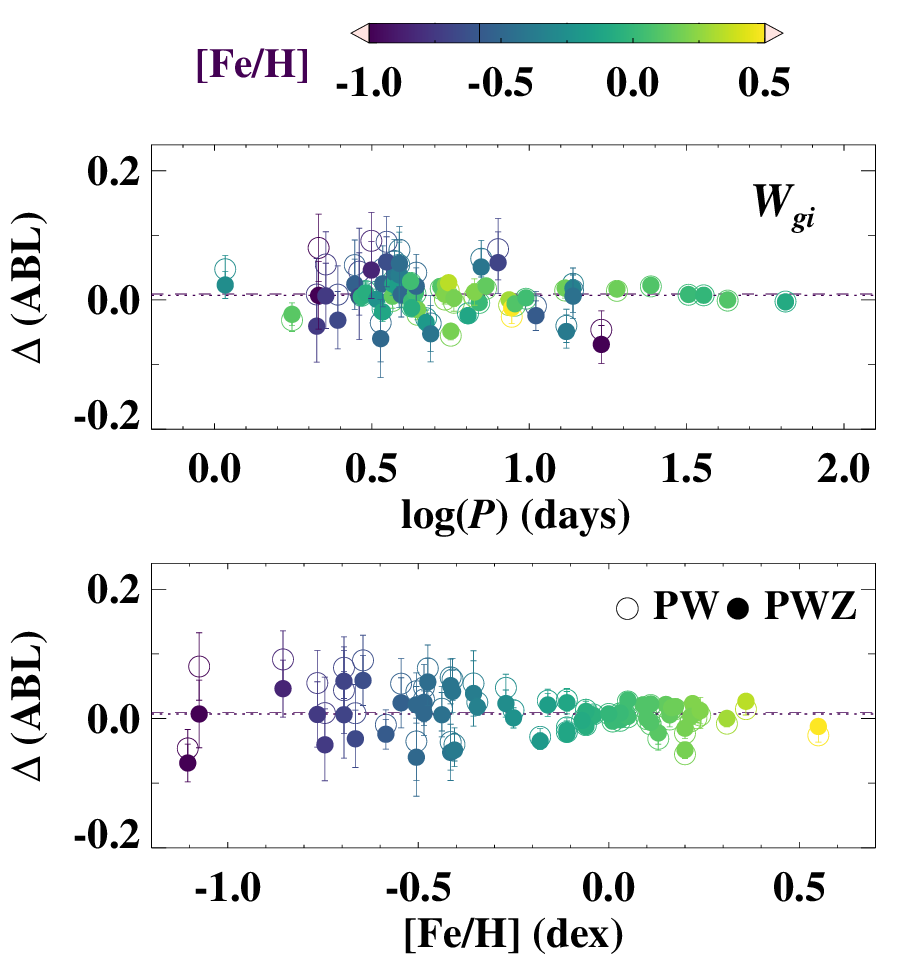} &  
      \includegraphics[width=0.3\textwidth]{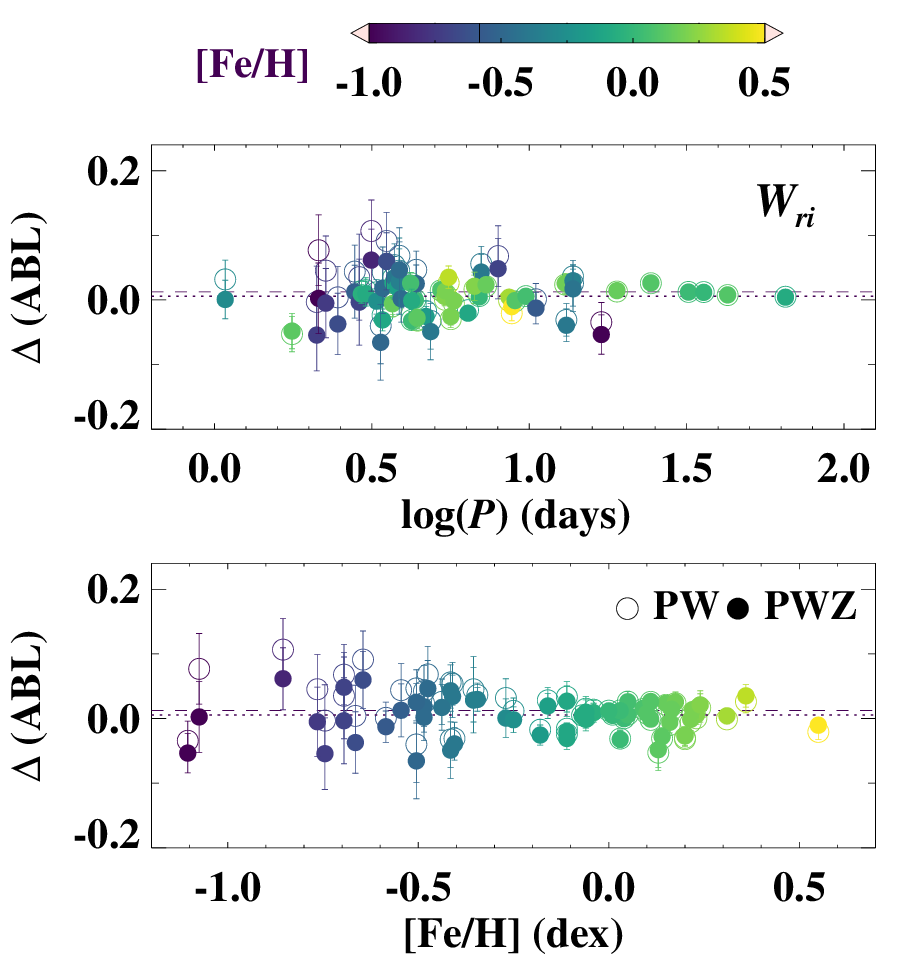} &  
      \includegraphics[width=0.3\textwidth]{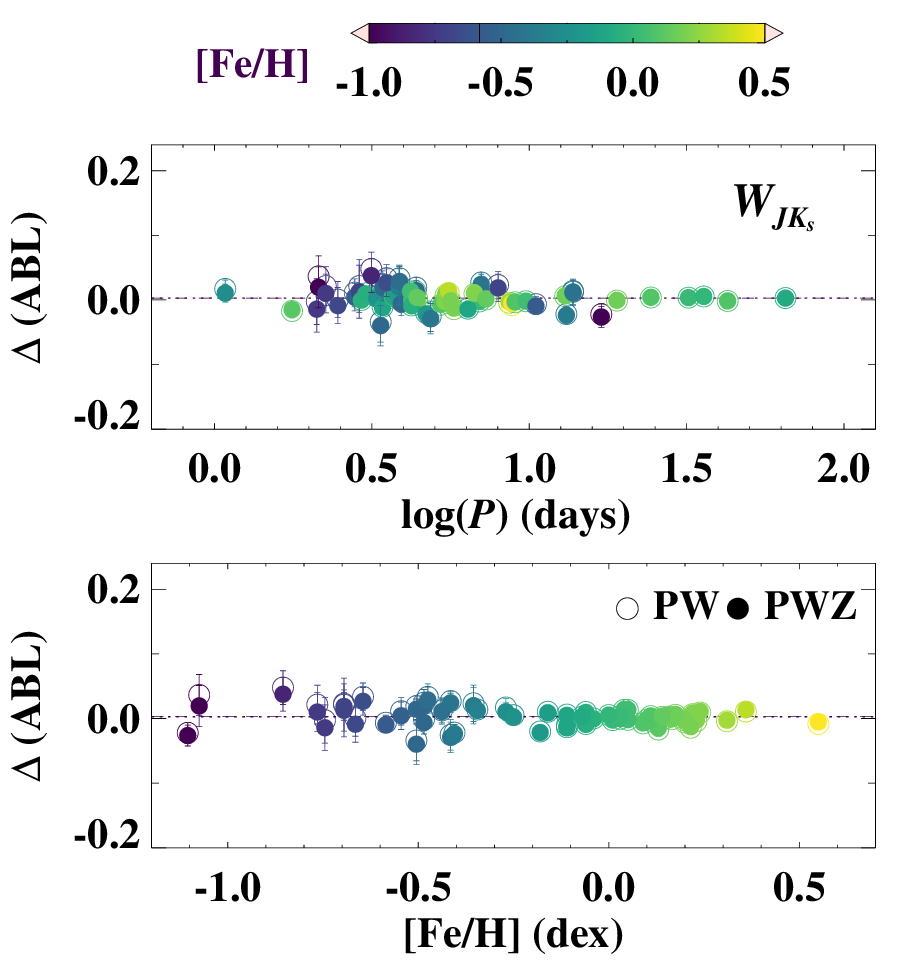} \\ 
\end{tabular}
        \caption{Comparison of residuals of $r, z, K_s, W_{gi}, W_{ri}, W_{JK_s}$ PL/PW and PLZ/PWZ relations is shown in six different panels. In each panel, the top and bottom plot shows the variation of residuals as a function of period and metallicity, respectively. The bandpasses are mentioned at the top right of each panel. The open and filled circles represent residuals of the PL/PW and PLZ/PWZ relations, respectively.
        The colour bar represents metallicity.}
             \label{fig:res_met}
\end{figure*}

\begin{figure*}
\centering
\includegraphics[width=0.85\textwidth]{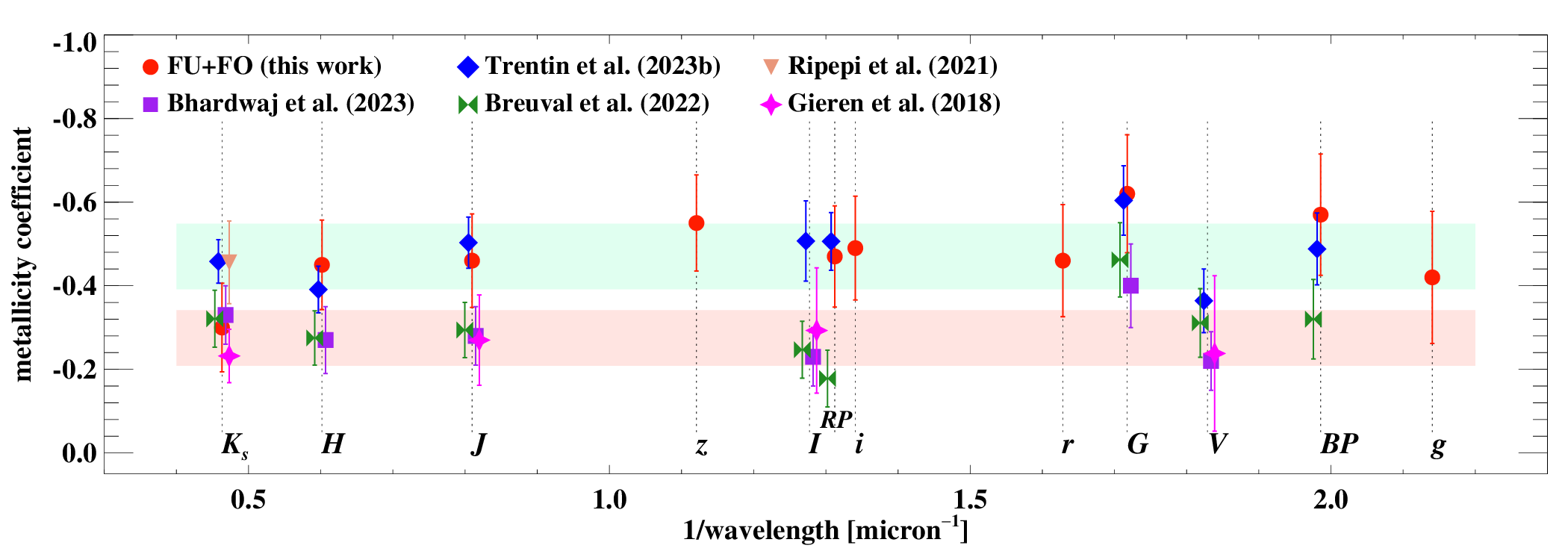}  
        \caption{Comparison of metallicity coefficients as a function of wavelength. The metallicity coefficients of the  $G, BP, RP, V, I, J, H,$ and  $K_s$ bands are taken from \citet{gieren2018, ripepi2021,
        breuval2022, trentin2023a, bhardwaj2023a}, but no previous measurements are available in the  $griz$ bands. The shaded regions display a $\pm1 \sigma$ standard
deviation around the mean value of the metallicity coefficients from (1) this work + \citet{trentin2023a} + \citet{ ripepi2021} and (2) \citet{bhardwaj2023a} + \citet{ breuval2022} + \citet{gieren2018}.}
        \label{fig:met_comp}
\end{figure*}

\begin{table*}
\begin{center}
\caption{Period-luminosity-metallicity and period-Wesenheit-metallicity relations for MW Cepheids. \label{tbl:plz}}
\begin{tabular}{cccccc}
\hline\hline
        {Band} & {$\alpha_\lambda$} & {$\beta_\lambda$} & {$\gamma_\lambda$} &{$\sigma_\textrm{ABL}$}& {$N$}\\
\hline
\multicolumn{6}{c}{All (FU+FO) Cepheids ($\log P_0=1.0$ days)}\\
\hline
              $G$ &$    -4.93\pm0.07    $&$    -3.30\pm0.10    $&$    -0.62\pm0.14    $&     0.04&  59\\
             $BP$ &$    -4.39\pm0.08    $&$    -2.88\pm0.11    $&$    -0.57\pm0.15    $&     0.04&  59\\
             $RP$ &$    -4.93\pm0.05    $&$    -2.84\pm0.08    $&$    -0.47\pm0.12    $&     0.03&  59\\
              $g$ &$    -4.24\pm0.07    $&$    -2.78\pm0.12    $&$    -0.42\pm0.16    $&     0.04&  58\\
              $r$ &$    -4.39\pm0.06    $&$    -2.62\pm0.10    $&$    -0.46\pm0.14    $&     0.04&  59\\
              $i$ &$    -4.59\pm0.05    $&$    -2.73\pm0.09    $&$    -0.49\pm0.13    $&     0.03&  59\\
              $z$ &$    -4.62\pm0.04    $&$    -2.73\pm0.08    $&$    -0.55\pm0.12    $&     0.03&  59\\
              $J$ &$    -5.64\pm0.04    $&$    -2.99\pm0.07    $&$    -0.46\pm0.12    $&     0.02&  59\\
              $H$ &$    -6.00\pm0.04    $&$    -3.23\pm0.07    $&$    -0.45\pm0.11    $&     0.02&  59\\
            $K_s$ &$    -6.05\pm0.04    $&$    -3.26\pm0.07    $&$    -0.30\pm0.11    $&     0.02&  58\\
\hline
            $W_G$ &$    -6.21\pm0.03    $&$    -3.54\pm0.06    $&$    -0.47\pm0.10    $&     0.02&  60\\
         $W_{gr}$ &$    -5.07\pm0.05    $&$    -3.10\pm0.10    $&$    -0.62\pm0.12    $&     0.04&  60\\
         $W_{ri}$ &$    -5.20\pm0.06    $&$    -3.36\pm0.09    $&$    -0.49\pm0.13    $&     0.03&  60\\
         $W_{gi}$ &$    -5.00\pm0.04    $&$    -2.93\pm0.07    $&$    -0.53\pm0.11    $&     0.03&  59\\
         $W_{iz}$ &$    -4.78\pm0.06    $&$    -2.87\pm0.10    $&$    -0.62\pm0.14    $&     0.04&  60\\
         $W_{JH}$ &$    -6.59\pm0.05    $&$    -3.60\pm0.09    $&$    -0.42\pm0.12    $&     0.01&  58\\
         $W_{iK}$ &$    -6.37\pm0.03    $&$    -3.35\pm0.06    $&$    -0.25\pm0.10    $&     0.01&  58\\
         $W_{JK}$ &$    -6.35\pm0.04    $&$    -3.42\pm0.07    $&$    -0.22\pm0.11    $&     0.01&  59\\
\hline
\end{tabular}
\end{center}
        \footnotesize{{\bf Notes:} The zero-point ($\alpha$), slope ($\beta$), metallicity coefficient ($\gamma$), dispersion ($\sigma_\textrm{ABL}$) of the ABL fits, and 
the number of stars ($N$) in the final PLZ/PWZ relations are listed.}
\end{table*}

The influence of metallicity on the absolute magnitudes of Cepheid variables needs to be properly quantified because it is a crucial parameter in improving the overall
fit of the cosmic distance ladder used to measure the Hubble constant \citep{riess2022a}. Most of the recent studies have measured a negative sign for the metallicity coefficient of PL relation 
\citep{gieren2018,riess2021,ripepi2021,breuval2022, bhardwaj2023a}. While our sample of stars with spectroscopic metallicities is rather small, it is based on a homogeneously collected high-resolution spectroscopic dataset within 
the framework of the C-MetaLL survey, thus minimizing systematics in combining different literature [Fe/H] measurements. 

The residuals of PL and PW relations in Figures \ref{fig:abl_plr} and \ref{fig:abl_pwr} show a clear trend as a function of metallicity such that metal-poor stars exhibit larger residuals. Figure~\ref{fig:hist_all}
displays metallicity distribution of 65 stars with available [Fe/H] values for our sample of Cepheids. The most metal-poor star ($\textrm{[Fe/H]}=-1.66$~dex) in our sample comes from {\it Gaia} and has
poor RVS quality flags. There are four more stars with similar quality flags and an assigned error of 0.5~dex \citep{trentin2023a}, which were excluded from this analysis. Therefore, our sample consists of 61 stars including
33 FU and 28 FO Cepheids. The [Fe/H] values of these Cepheids are between $-1.11$ and $0.6$~dex. Given small samples of both subtypes, we only derive PLZ and PWZ relations for the combined sample of Cepheid variables. A metallicity term is added to equation~(\ref{eq:plr_abl})
in the following form:

\begin{eqnarray}
 {\textrm{ABL}} ~&=&~ \overline{\omega}_\textrm{(mas)}10^{0.2m_\lambda-2} \nonumber \\
        ~&=&~ 10^{0.2(\alpha_{\lambda} + \beta_{\lambda} (\log P_{i} - \log P_0) + \gamma_{\lambda}\textrm{[Fe/H]})}.
    \label{eq:plz_abl}
\end{eqnarray}

In order to better constrain the metallicity coefficients, we assume that the slope of the PL relation does not change with metallicity. This is a basic assumption that is applied when measuring 
extragalactic distances using Cepheid variables. However, the metallicity can also affect the slope of the PL relations at different wavelengths \citep[e.g.][]{ripepi2021, trentin2023a}. Given a small sample size, we assume a fixed slope of PL/PW relation from Tables~\ref{tbl:plrs} and \ref{tbl:pwrs} in equation~(\ref{eq:plz_abl}) at a given wavelength.
Therefore, we only solve for the absolute zero-point and the metallicity coefficient as free parameters. 

Table~\ref{tbl:plz} lists the coefficients of PLZ and PWZ relations. The metallicity coefficient varies significantly between $-0.62\pm0.14$ mag/dex in $G$ band and $-0.30\pm0.11$ mag/dex in 
$K_s$ band. These metallicity coefficients are systematically larger than those determined by \citet{breuval2022} and \citet{bhardwaj2023a}, but are more in agreement with previous results from the C-MetaLL survey \citep{ripepi2021, trentin2023a}. We note that the metallicity range of MW Cepheid sample was rather small in \citet{breuval2022} and \citet{bhardwaj2023a} with a mean value of $0.09$~dex (dispersion of $0.12$ dex) and $0.04$~dex (dispersion of $0.09$ dex), respectively. In contrast, the mean metallicity of the sample used in this work is $-0.18$ dex (median value of $-0.11$ dex) and a three times larger dispersion of $0.37$ dex than \citet{breuval2022} sample. If we restrict the sample to 38 stars with 
[Fe/H] values larger than $-0.3$~dex, the metallicity coefficients decrease (in absolute sense) considerably but exhibit larger uncertainties. For example, the metallicity coefficient in $r$-band becomes $-0.20\pm0.24$ mag/dex from a value of $-0.46\pm0.14$ mag/dex quoted in Table~\ref{tbl:plz}. These coefficients are almost zero in the PLZ/PWZ relations involving the $K_s$ band when most metal-poor stars ([Fe/H]$<-0.3$~dex) are excluded. While the metallicity coefficients listed in Table~\ref{tbl:plz} are weakly
constrained due to lower statistics, it is important to probe this dependence more in detail. These large metallicity coefficients of the order of $-0.5$ mag/dex can lead to significant biases
of $0.05$ mag in distance modulus determinations if the mean metallicities of the calibrator and target Cepheid PL relations differ by even $\Delta\textrm{[Fe/H]}=0.1$ dex.

\subsection{Comparison of residuals of PL and PLZ relations}
Figure~\ref{fig:res_met} displays the residuals of PL/PW and PLZ/PWZ relations as a function of metallicity. It is clearly evident that the dominant contribution to the metallicity term comes
from the metal-poor stars ($\textrm{[Fe/H]}<-0.3$~dex) which also have larger parallax uncertainties. The median parallax uncertainties for stars with $\textrm{[Fe/H]}>-0.3$~dex is $6.4\%$ while it increases to $16.7\%$
for Cepheids with $\textrm{[Fe/H]}<-0.3$~dex. For example, the median residuals of $r$ band PL relation are $\Delta(\textrm{ABL})=0.004$ and $\Delta(\textrm{ABL})=0.048$ for metallicities higher 
(metal-rich) and lower (metal-poor) than $-0.3$~dex, respectively. When the metallicity term is included, the median residual of metal-poor stars decreases to $\Delta(\textrm{ABL})=0.014$, 
while it remains the same for metal-rich stars. Similarly, the median residuals of the ABL fits for $W_{JK_s}$ Wesenheit relations decrease from $\Delta(\textrm{ABL})=0.021$ to 
$\Delta(\textrm{ABL})=0.012$ for metal-poor stars. However, this is also expected since the median metallicities for metal-rich and metal-poor stars are $0.10$ and $-0.51$~dex, respectively.
Therefore, the contribution to average residuals due to metallicity coefficients is smaller for metal-rich stars than for metal-poor stars. Nevertheless, it is difficult to separate the
contribution of metallicity effects and parallax uncertainties to the scatter in the residuals seen for metal-poor stars in Figure~\ref{fig:res_met}. If we exclude stars with larger parallax
uncertainties, there are not enough metal-poor Cepheids in the sample for a proper quantification of metallicity coefficient of PL/PW relations. 

\subsection{Metallicity coefficient as a function of wavelength}

We compared the metallicity coefficients of PLZ/PWZ relations with recent determinations in the literature. Figure~\ref{fig:met_comp} displays metallicity coefficient as a function of wavelengths.
\citet{breuval2022} fitted a linear regression between $\gamma$ and $1/\lambda$ and concluded that the metallicity effect is uniform over a wide range of wavelength. We do not see a strong linear
correlation between metallicity coefficient and wavelength. The PLZ relations at longer wavelengths seem to suggest a marginally smaller metallicity term than at shorter wavelengths, but the
uncertainties on $\gamma$ values are larger for our sample of stars. The metallicity coefficients derived in this work are consistent with earlier results of \citet{ripepi2021} and \citet{trentin2023a} within the 
C-MetaLL survey. We note that the approach of using individual metallicities of MW Cepheids is the same in these studies. In contrast, the results of \citet{gieren2018}, \citet{breuval2022}, and \citet{bhardwaj2023a} are based on a comparison of intercepts of PL relations for Cepheids in the MW and the Magellanic Clouds. The average value of the metallicity coefficient ($-0.28\pm0.07$ mag/dex) derived in these studies is systematically smaller than those listed
in Table~\ref{tbl:plz}. In contrast, the mean value of the metallicity coefficient ($-0.47\pm0.08$~mag/dex) derived from all the values in this work, and \citet{ripepi2021}, and \citet{trentin2023a} is significantly larger. However, these metallicity coefficients are also based on a sample of Cepheids covering a wide range of metallicities. Nevertheless, the metallicity coefficients of our PLZ/PWZ relations are in agreement with most of these measurements given their large uncertainties due to a small sample of
Cepheid variables.

\subsection{Parallax correction and the uncertainty in the zero-point}

As shown in the previous subsections, the sample presented in this work is not optimal for the accurate and precise calibration of PLZ and PWZ relations due to low statistics, large parallax uncertainties, and the lack of accurate reddening measurements. 
Nevertheless, we investigate the reliability of the large metallicity coefficients and the zero points of the calibrated PLZ and PWZ relations.
When compared with the results of \citet{bhardwaj2023a}, the zero-points of $K_s$ band PLZ, and $W_{J,Ks}/W_G$ PWZ relations are nearly $\sim0.25$ mag brighter in this work. We note that both studies adopted the same parallax zero-point offset correction of $-14\mu$as \citep{riess2021}, but \citet{bhardwaj2023a} sample is based on MW Cepheid standards that have accurate parallaxes, low reddening, and a limited metallicity range. Therefore, the readers interested in determining Cepheid-based distances are recommended to use the calibrated relations provided in \citet{bhardwaj2023a} for metallicities closer to solar value, and those in \citet{trentin2023a} for a wide-range of metallicities.

In the recent work within the C-MetaLL survey, \citet{trentin2023a} suggested that a larger parallax zero-point offset correction results in a larger distance measurement. The authors found that both the slope and the intercept of PLZ/PWZ relation increases by $1-4\%$ and the metallicity coefficient decreases (in absolute sense) if a larger parallax offset correction is adopted. For the $W_{J,K_s}$ Wesenheit, \citet[][Table 3]{trentin2023a} found the coefficients of PWZ relation: $\alpha=-6.09\pm0.02$ mag, $\beta=-3.29\pm0.03$ mag/dex, $\gamma=-0.45\pm0.05$ mag/dex, with no parallax zero-point offset correction. When compared with the results in Table~\ref{tbl:plz} based on a parallax zero-point offset correction of $-14~\mu$as, indeed the slope and the intercept increase and the metallicity coefficient decreases (in absolute sense), in agreement with \citet{trentin2023a}. However, the zero-point is significantly brighter by $\sim 0.25$ mag, but this difference becomes smaller when no parallax zero-point offset correction is adopted. For this latter case, we find the coefficients of $W_{J,K_s}$ PWZ relation: $\alpha=-6.25\pm0.04$ mag, $\beta=-3.42\pm0.07$ mag/dex, $\gamma=-0.43\pm0.10$ mag/dex. We note the slope is fixed in our analysis, and the metallicity coefficient is now in excellent agreement with \citet{trentin2023a}. However, the zero-point is still significantly brighter by $\sim 0.16$ mag, due to larger parallax uncertainties. In terms of the distance to the LMC, \citet[][Table 3]{trentin2023a} found a value of $18.41\pm0.02$ mag using $W_{J,K_s}$ Wesenheit. For the LMC Cepheids, we adopt the zero-point of $12.46\pm0.05$~mag for $W_{J,Ks}$ Wesenheit from \citet{bhardwaj2016a} based on the data from \citet{macri2015}, and a mean-metallicity of $-0.41\pm0.02$ \citep{romaniello2022}. We find a LMC distance of $18.53\pm0.06$~mag for the no parallax zero-point offset correction. However this value of LMC distance increases to $18.72\pm0.06$~mag for the calibrated PWZ relation listed in Table~\ref{tbl:plz} due to a smaller metallicity term and a brighter absolute zero-point. With the sample presented in this work, it is not possible to separate the contribution of parallax uncertainties (and parallax zero-point offset correction) and metallicity effects on the absolute calibration of PLZ and PWZ relations.

\section{Summary}
\label{sec:discuss}

We presented new homogeneously collected light curves of 78 MW Cepheid variables at optical Sloan ($griz$) and NIR ($JHK_s$) wavelengths. These observations were obtained simultaneously using the
REM telescope and present the light curves in $z$ band for FU Cepheids and in $griz$ band for FO Cepheids for the first time. Multiband light curves of Cepheids were obtained primarily to 
complement the high-resolution spectroscopic metallicities within the framework of the C-MetaLL survey \citep{ripepi2021}. The sample of 78 Cepheids includes 49 FU and 29 FO mode variables. The light curves of Cepheids
were fitted with Fourier sine series to determine accurate mean magnitude and peak-to-peak amplitudes, which were used to investigate their pulsation properties. The period-amplitude
diagrams were presented for the time in Sloan filters for both subtypes of Cepheids. The mean magnitudes were used to derive PL and PW relations for Cepheid variables using their astrometry-based
luminosity at multiple wavelengths. 

Cepheid variables in the present study are located at a wide range of distances between 0.5 and 19.7 kpc, and therefore their {\it Gaia} parallaxes exhibit varying uncertainties with larger errors
for distant objects. Moreover, some of these Cepheids are located in the Galactic disk and  anti-centre direction and are significantly reddened with colour-excess values exceeding 1 mag. 
There are no reddening values available in the literature for most of these Cepheids, and thus the colour-excess values were obtained by determining their intrinsic colours using the empirical period-colour
relations for Cepheid variables \citep{tammann2003, ripepi2021}. The larger uncertainties in parallaxes and reddening values and a modest sample size limits the accuracy and precision of 
the PL relations for these variables. The reddening uncertainties can be mitigated by employing the Wesenheit magnitudes, which results in  tighter PW relations, in particular, 
at shorter optical wavelengths where extinction is more severe.

In addition to parallax uncertainties, metallicity variations can also contribute to the scatter in the empirical PL and PW relations. Homogeneous high-resolution spectroscopic metallicities for 59 of
78 Cepheids have already been collected as part of the C-MetaLL survey. In addition, six Cepheids have medium-resolution  spectra from {\it Gaia}-RVS \citep{blanco2022},   two of which  have reliable quality flags. 
We investigated the residuals of astrometry-based luminosity with PL/PW relation fits as a function of their metallicities, if available. The residuals exhibit a clear trend with metallicity such 
that metal-poor stars have higher astrometry-based luminosity and fainter absolute magnitudes. This trend becomes significant for $\textrm{[Fe/H]}<-0.3$~dex, where the parallax uncertainties are also larger. 
When deriving PLZ and PWZ relations, the metallicity coefficient of these relations varies between $-0.30\pm0.11$ mag/dex in $K_s$ band to $-0.55\pm0.12$ mag/dex in $z$ band. While these empirical metallicity coefficients
are weakly constrained due to the  small sample size, they are systematically larger than previous determinations at optical and NIR filters. We compared the residuals of PL/PW
and PLZ/PWZ relations to further investigate the impact of including the metallicity term, and found that the metallicity contribution predominantly affects the residuals for the most metal-poor Cepheids ($\textrm{[Fe/H]}<-0.3$~dex). The metallicity effect becomes smaller if we exclude these metal-poor Cepheids from the sample, but the uncertainties on the metallicity coefficient increase due to low statistics.
The small sample size prevents us from separating the contribution to the scatter in the PL/PW relations due to the metallicity term and parallax uncertainties. These large metallicity coefficients and bright zero-points should be treated cautiously, and a larger sample of metal-poor Cepheids with 
more accurate parallaxes is needed to confirm these relatively large metallicity coefficients of the PLZ/PWZ relations. 

The aim of the ongoing C-MetaLL survey is to increase the sample of Cepheids with homogeneous photometric and spectroscopic data to a few hundred  stars, which is ideal for the absolute calibration of the 
Leavitt law and a proper quantification of metallicity effects at multiple wavelengths. In addition, the extension of the photometric data presented in this paper will also be useful for several other scientific
goals, for example  investigating the light curve structure of these variables at multiple wavelengths. The simultaneous optical and NIR light curves will be useful for a quantitative comparison with
the theoretically predicted light curves, thus providing strong constraints for the input parameters to the pulsation models. Multiwavelength data will also be used to constrain the reddening values for
Cepheid variables in a future study. Moreover, light curves in the optical Sloan filters will serve as templates for
the identification and classification of Cepheid variables in the Vera C. Rubin Observatory's Legacy Survey of Space and Time.

\begin{acknowledgements}

We thank the anonymous referee for the quick and constructive referee report that helped improve the manuscript. This project has received funding from the European Union’s Horizon 2020 research and innovation programme under the Marie Skłodowska-Curie grant agreement No. 886298.  This research was supported by the Munich Institute for Astro-, Particle and BioPhysics (MIAPbP) which is funded by the Deutsche Forschungsgemeinschaft (DFG, German Research Foundation) under Germany´s Excellence Strategy - EXC-2094 - 390783311. We acknowledge funding from INAF GO-GTO grant 2023 ``C-MetaLL - Cepheid metallicity in the Leavitt law'' (P.I. V. Ripepi). This work has made use of data from the European Space Agency (ESA) mission {\it Gaia} (\url{https://www.cosmos.esa.int/gaia}), processed by the {\it Gaia} Data Processing and Analysis Consortium (DPAC, \url{https://www.cosmos.esa.int/web/gaia/dpac/consortium}). Funding for the DPAC has been provided by national institutions, in particular the institutions participating in the {\it Gaia} Multilateral Agreement.
\end{acknowledgements}

%
%
\bibliographystyle{aa}
\bibliography{mybib_final.bib}

\end{document}